\documentclass[prd,showpacs,amsmath,amssymb,floatfix]{revtex4}
\usepackage{graphicx,color,dcolumn,booktabs}
\usepackage{longtable,lscape}

\begin{document}
\title{Newly observed $D_{sJ}(3040)$ and the radial excitations of P-wave charmed-strange mesons}

\author{Zhi-Feng Sun}
\author{Xiang Liu\footnote{Corresponding author}}\email{xiangliu@lzu.edu.cn}
\affiliation{School of Physical Science and Technology, Lanzhou University, Lanzhou 730000,  China}

\date{\today}
\begin{abstract}
Inspired by the newly observed $D_{sJ}(3040)^+$ state, in this work we systemically study the two-body strong decays of P-wave charmed-strange mesons with the first radial excitation. Under the assignment of $1^{+}(j^P=\frac{1}{2}^+)$, i.e. the first radial excitation of $D_{s1}(2460)^+$, we find that the width of $D_{sJ}(3040)^+$ is close to the lower limit of the BaBar measurement. This indicates that it is reasonable to interpret $D_{sJ}(3040)^+$ as the first radial excitation of $D_{s1}(2460)^+$. Our calculation further predicts that $0^-+1^-$ channels e.g. $D^+K^{*0}$, $D^0 K^{*+}$ and $D_s^+\phi$ are important for the search for $D_{sJ}(3040)^+$. To help future experiments finding the remaining three P-wave charmed-strange mesons with the first radial excitation, we present the predictions for the strong decays of these three P-wave charmed-strange mesons.
\end{abstract}

\maketitle

\section{Introduction}\label{sec1}

With the new observation of $D_{sJ}$ meson, the spectrum of the charmed-strange state is becoming abundant. So far, there exist six established charmed-strange mesons $D_s(1968)$, $D_s^*(2112)$, $D^*_{s0}(2317)$, $D_{s1}(2460)$, $D_{s1}(2536)$, $D_{s2}^*(2573)$ listed in Particle Data Group (PDG) \cite{Amsler:2008zzb}, which can be categorized as three doublets in terms of the heavy quark limit: $H=(0^-,1^-)=(D_s(1968), D_s^*(2112))$, $S=(0^+,1^+)=(D^*_{s0}(2317), D_{s1}(2460))$ and $T=(1^+,2^+)=(D_{s1}(2536),D_{s2}^*(2573))$.
Two years ago, a new charmed-strange meson $D_{s1}^*(2710)$ with $J^{P}=1^-$ was firstly announced by the Babar Collaboration \cite{Aubert:2006mh} and confirmed by the Belle Collaboration later \cite{:2007aa}. Very recently the Babar experiment found $D_{s1}^*(2710)$ again in the $D^*K$ invariant mass spectrum \cite{Aubert:2009di}. Another newly observed charmed-strange meson is $D_{sJ}^*(2860)$, which was observed in both $DK$ \cite{Aubert:2006mh} and $D^*K$ channels \cite{Aubert:2009di}. The phenomenological proposals of the quantum number of $D_{sJ}^*(2860)$ include $J^P=3^-$ \cite{Colangelo:2006rq,Zhang:2006yj} and $J^{P}=0^+$ \cite{vanBeveren:2006st,Close:2006gr,Zhang:2006yj}. As indicated by the Babar experiment, $J^P=0^+$
assignment for $D_{sJ}^*(2860)$ is forbidden according to the parity conservation since the $D^*K$ decay mode of $D_{sJ}^*(2860)$ was observed in Ref. \cite{Aubert:2009di}. A series of theoretical work \cite{Colangelo:2006rq,Zhang:2006yj,vanBeveren:2006st,Close:2006gr,
Wei:2006wa,Colangelo:2007ds,Vijande:2008zn,Zhang:2009nu,vanBeveren:2009jq} relevant to $D_{s1}^*(2710)$ and $D_{sJ}^*(2860)$ were carried out.

\begin{center}
\begin{figure}[htb]
\begin{tabular}{c}
\scalebox{1.2}{\includegraphics{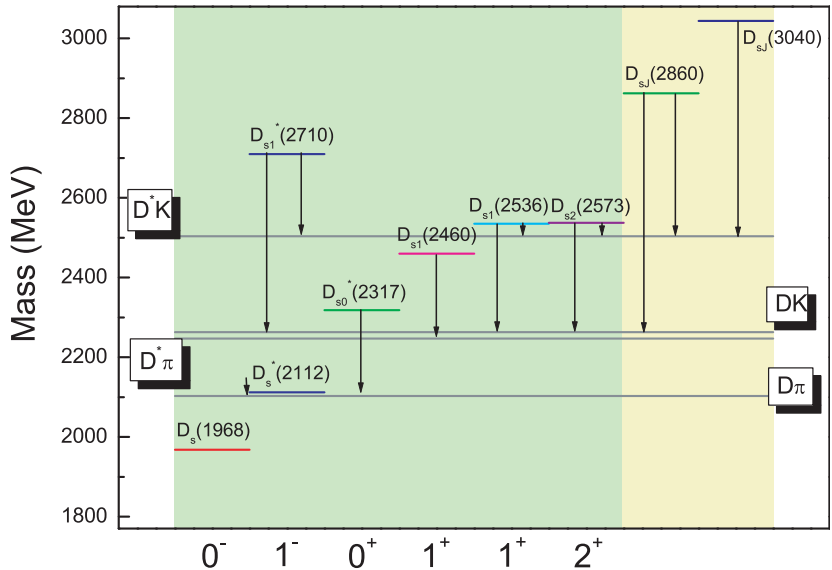}}
\end{tabular}
\caption{The mass spectrum of the observed charmed-strange mesons and the corresponding strong decay modes observed in experiment \cite{Aubert:2006mh,:2007aa,Aubert:2009di,Amsler:2008zzb}. \label{mass}}
\end{figure}
\end{center}

Besides the observations of $D_{s1}^*(2710)$ and $D_{sJ}^*(2860)$ by analyzing the $D^*K$ invariant mass spectrum in inclusive $e^+e^-$ interactions \cite{Aubert:2009di}, Babar also announced a new charmed-strange state $D_{sJ}(3040)$ with the mass $M=3044\pm8(\mathrm{stat})^{+30}_{-5}(\mathrm{syst})$ MeV and the width $\Gamma=239\pm35(\mathrm{stat})^{+46}_{-42}(\mathrm{syst})$ MeV \cite{Aubert:2009di}.
The observation of $D_{sJ}(3040)$ not only makes the spectrum of the charmed-strange meson abundant (the mass spectrum of the observed charmed-strange mesons are listed in Fig. \ref{mass}), but also stimulates our interest in exploring its underlying structure.

As indicated by the Babar Collaboration, $D_{sJ}(3040)$ was only observed in $D^*K$ channel while not find in $DK$ decay mode. Thus, its possible quantum number includes $J^{P}=1^+,\,0^-,\, 2^-, \cdots$. Since $D_{s1}(2710)(J^P=1^-)$ is the first radial excitation of $D_{s}^*(2112)$ and the mass of $D_{sJ}(3040)$ is far larger than that of $D_{s1}(2710)$, thus we further exclude $0^-$ assignment, \textit{viz}. the first radial excitation of $D_s(1968)$ for $D_{sJ}(3040)$. In Ref. \cite{Matsuki:2006rz}, Matsuki, Morii and Sudoh once predicted the mass of $c\bar{s}$ state with $n^{2s+1}L_J=2^{3}P_1$: $m=3082$ MeV, which is close to the experimental value of the mass of $D_{sJ}(3040)$. Thus, $1^+$ assignment to $D_{sJ}(3040)$, the first radial excitation of $D_{s1}(2460)$, becomes the most possible.

If $D_{sJ}(3040)$ as the radial excitation of P-wave charmed-strange state is true, further experiment is of the potential to search the rest three radial excitations of P-wave charmed-strange states. Thus, a systematical phenomenological study of the strong decay mode of P-wave charmed-strange mesons with the first radial excitation is an important and interesting topic. By this study, we will not only obtain the information the decays of these P-wave charmed-strange mesons, but also can test $1^+$ quantum number assignment to $D_{sJ}(3040)$ comparing calculated decay width with the experimental data.

In this work, we will be dedicated to the study of the strong decay modes of P-wave charmed-strange mesons with the radial excitation by the $^3P_0$ model \cite{Micu:1968mk,yaouanc,LeYaouanc:1977gm,LeYaouanc:1988fx,vanBeveren:1979bd,Bonnaz:2001aj,sb}. Further we will obtain the information of the order of magnitude of the strong decay modes of $D_{sJ}(3040)$.

The paper is organized as follows. After the introduction, we
briefly review the $^3P_0$ model and present the formulation of the strong decays of P-wave charmed-strange mesons with the radial excitation. Finally, the numerical result will be shown. The last section is a short summary.

\section{The strong decay of P-wave charmed-strange mesons with the radial excitation}\label{sec2}

Before illustrating the strong decay of P-wave charmed-strange mesons with the radial excitation, we first introduce the category of the heavy flavor meson.

In the heavy quark limit $m_{Q}\to \infty$, the heavy quark plays a role of a static color source to interact with the light part within the heavy flavor hadron. Thus, the spin of the heavy quark $\vec{s}_{Q }$ can be separated from the total angular momentum $J$ of the heavy flavor hadron. Furthermore, $\vec{j}_{\ell}=\vec{s}_{q}+\vec{L}$ is a good quantum number, where $\vec{s}_{q}$ and $\vec{L}$ denote the spin of the light part of the heavy flavor hadron and the orbital angular momentum between the light part and the heavy quark, respectively.

Thus, the heavy mesons can be grouped into doublets according to
$j_{\ell}^P$, which include $j_{\ell}=\frac{1}{2}^-$ doublet $(0^-, 1^-)$ with the orbital angular
momentum $L=0$, $j_{\ell}=\frac{1}{2}^+$ doublet $(0^+,1^+)$ and
$\frac{3}{2}^+$ doublet $(1^+,2^+)$ with $L=1$. For $L=2$ there exist
$(1^{-},2^{-})$ and $(2^{-},3^{-})$ doublets with
$j_{\ell}^P=\frac{3}{2}^-$ and $\frac{5}{2}^-$, respectively. As showing in Fig. \ref{mass}, the states existing in the doublets $(0^-,1^-)$, $(0^+,1^+)$ and $(1^+,2^+)$ are already filled with the observed charmed-strange mesons. Two $1^+$ states existing $S$ and $T$ are the mixture of two basis states $1^1P_1$ and $1^3P_1$ \cite{Lu:2006ry,Luo:2009wu}
\begin{eqnarray}
\left (\begin{array}{c}
 \left|1^+,j_l^P=\frac{1}{2}^+\right\rangle\nonumber\\
\left|1^+,j_l^P=\frac{3}{2}^+\right\rangle
\end{array}
\right )=\left(\begin{array}{cc} \cos\theta&\sin\theta\\
                                 -\sin\theta&\cos\theta\end{array}\right)
\left (\begin{array}{c}
 \left|1^1P_1\right\rangle\nonumber\\
\left|1^3P_1\right\rangle
\end{array}
\right )\,,
\end{eqnarray}
where one takes the mixing angle
$\theta=-\tan^{-1}\sqrt{2}=-54.7^\circ$ according to the estimate in
the heavy quark limit.

For P-wave charmed-strange mesons with the radial excitation discussed in this work, one also categorizes them as $S=(0^+,\,1^+)$ and $T=(1^+,\,2^+)$ doublets according to the above approach.
Two $1^+$ states are the mixture of two basis states $2^1P_1$ and $2^3P_1$, which satisfy the below relation
\begin{eqnarray}
\left (\begin{array}{c}
 \left|1^+,j_l^P=\frac{1}{2}^+\right\rangle\nonumber\\
\left|1^+,j_l^P=\frac{3}{2}^+\right\rangle
\end{array}
\right )=\left(\begin{array}{cc} \cos\theta'&\sin\theta'\\
                                 -\sin\theta'&\cos\theta'\end{array}\right)
\left (\begin{array}{c}
 \left|2^1P_1\right\rangle\nonumber\\
\left|2^3P_1\right\rangle
\end{array}
\right )\,.
\end{eqnarray}
In this work, we approximately take $\theta'=\theta=-54.7^\circ$.

For distinguishing P-wave states with and without the first radial excitation, one labels four P-wave states without the first radial excitation as $0^{+}(S)$, $1^{+}(S)$, $1^{+}(T)$
and $2^{+}(T)$. Four P-wave states with the first radial excitation are named as $0^{+}(S^\star)$, $1^{+}(S^\star)$, $1^{+}(T^\star)$
and $2^{+}(T^\star)$.

If we set the upper limit of the masses of P-wave states with the first radial excitation as 3.04 GeV, the kinematically allowed decay modes of P-wave states with the first radial excitation are listed in Table \ref{decay modes}.
In the following, the $^3P_0$ model will be applied to calculate these strong decays in Table \ref{decay modes}.

\renewcommand{\arraystretch}{1.3}
\begin{table}[htb]
\begin{tabular}{c|c|c}\toprule[1pt]
State&Decay modes& Decay channels\\\midrule[1pt]
&$0^-+0^-$&$D^+K^0,\,D^0K^+,\,D_{s}^+\eta^{(\prime)}$\\
&$1^-+1^-$&$D^{*+}K^{*0},\,D^{*0} K^{*+}$\\
&$0^++1^-$&$\times$\\

$0^+(S^\star)$&$1^+(S)+0^-$&$D_1(2430)^0K^+,\,D_1(2430)^+K^0,\,D_{s1}(2460)^{+}\eta$\\
&$1^+(S)+1^-$&$\times$\\
&$1^+(T)+0^-$&$D_1(2420)^+K^0,\,D_1(2420)^0K^+$\\
&$1^+(T)+1^-$&$\times$\\
&$2^++1^-$&$\times$\\\midrule[1pt]

&$0^-+1^-$&$D^+K^{*0},\,D^0 K^{*+},\,D_s^+\phi$\\&$1^-+0^-$&$D^{*+}K^{0},\,D^{*0} K^{+},\,D_{s}^{*+}\eta$\\
&$1^-+1^-$&$D^{*+}K^{*0},\,D^{*0} K^{*+}$\\&$0^++0^-$&$D^*_0(2400)^{+}K^{0},\,D_0^*(2400)^{0} K^{+},\,D_{s0}^{*}(2317)^+\eta$\\
$1^+(S^\star)/1^+(T^\star)$&$0^++1^-$&$\times$\\&$1^+(S)+0^-$&$D_1(2430)^0K^+,\,D_1(2430)^+K^0,\,D_{s1}(2460)^{+}\eta$\\
&$1^+(T)+0^-$&$D_1(2420)^+K^0,\,D_1(2420)^0K^+$\\
&$1^+(S)+1^-$&$\times$\\&$1^+(T)+1^-$&$\times$\\
&$2^++0^-$&$D_{2}^*(2460)^+K^0,\,D_2^*(2460)^0K^+$\\
&$2^++1^-$&$\times$\\ \midrule[1pt]

&$0^-+0^-$&$D^+K^0,\,D^0K^+,\,D_{s}^+\eta^{(\prime)}$\\

&$0^-+1^-$&$D^+K^{*0},\,D^0 K^{*+},\,D_s^+\phi$\\&$1^-+0^-$&$D^{*+}K^{0},\,D^{*0} K^{+},\,D_{s}^{*+}\eta$\\
&$1^-+1^-$&$D^{*+}K^{*0},\,D^{*0} K^{*+}$\\
&$0^++1^-$&$\times$\\
$2^+(T^\star)$&$1^+(S)+0^-$&$D_1(2430)^0K^+,\,D_1(2430)^+K^0,\,D_{s1}(2460)^{+}\eta$\\
&$1^+(S)+1^-$&$\times$\\
&$1^+(T)+0^-$&$D_1(2420)^+K^0,\,D_1(2420)^0K^+$\\
&$1^+(T)+1^-$&$\times$\\
&$2^++0^-$&$D_{2}^*(2460)^+K^0,\,D_2^*(2460)^0K^+$\\
&$2^++1^-$&$\times$\\
\bottomrule[1pt]
\end{tabular}\caption{The relevant strong decay modes of P-wave charmed-strange mesons with the first radial excitation allowed by the conservation of the quantum number. Here ``$\times$" denotes
that the decay modes are kinematically forbidden if setting the upper limit of the masses of P-wave states with the first radial excitation as 3.04 GeV. Since the
$1^+$ state in the $(1^+, 2^+)$ doublet decays into $D^\ast \pi$
via D-wave, it is very narrow and denoted as $D_1(2420)$
\cite{Amsler:2008zzb}. The $1^+$ state in the $(0^+, 1^+)$ doublet decays
into $D^\ast \pi$ via S-wave. Hence, it is very broad and denoted as
$D_1(2430)$ \cite{Amsler:2008zzb}.
\label{decay modes}}
\end{table}

\subsection{A review of the QPC model}

The $^3P_0$ model, also
known as the Quark Pair Creation (QPC) model, was firstly proposed by Micu in Ref. \cite{Micu:1968mk} to calculate Okubo-Zweig-Iizuka (OZI) allowed strong decays of a meson. Later, this model was developed by the other theoretical groups \cite{yaouanc,LeYaouanc:1977gm,LeYaouanc:1988fx,vanBeveren:1979bd,Bonnaz:2001aj,sb} and is successful when applied
extensively to the calculation of the strong decay of hadron
\cite{Blundell:1995ev,Page:1995rh,Capstick:1986bm,Capstick:1993kb,Ackleh:1996yt,Zhou:2004mw,Guo:2005cs,Close:2005se,Lu:2006ry,Zhang:2006yj,Chen:2007xf,Li:2008mz,Luo:2009wu}.


In the QPC model, the heavy flavor meson decay occurs through a
quark-antiquark pair production from the vacuum, which is of the quantum
number of the vacuum, i.e. $0^{++}$ \cite{Micu:1968mk,yaouanc}. For describing a strong decay process of the charmed-strange meson $A(c(1)\bar{s}(2))\to B(c(1)\bar{q}(3))+C(\bar{s}(2)q(4))$, one writes out the S-matrix \begin{eqnarray}
\langle
BC|S|A\rangle=I-i2\pi\delta(E_f-E_i)\langle{}BC|T|A\rangle.\label{smatrix}
\end{eqnarray}
In the non-relativistic limit, the transition operator $T$ is depicted as
\begin{eqnarray}
T&=& - 3 \gamma \sum_m\: \langle 1\;m;1\;-m|0\;0 \rangle\,
\int\!{\rm d}{\textbf{k}}_3\; {\rm
d}{\textbf{k}}_4 \delta^3({\textbf{k}}_3+{\textbf{k}}_4) \nonumber\\&&\times{\cal
Y}_{1m}\left(\frac{{\textbf{k}}_3-{\textbf{k}_4}}{2}\right)\;
\chi^{3 4}_{1, -\!m}\; \varphi^{3 4}_0\;\,
\omega^{3 4}_0\; d^\dagger_{3i}({\textbf{k}}_3)\;
b^\dagger_{4j}({\textbf{k}}_4)\,,\label{tmatrix}
\end{eqnarray}
where $i$ and $j$ denote the $SU(3)$ color indices of the created quark
and anti-quark. $\varphi^{34}_{0}=(u\bar u +d\bar d +s \bar
s)/\sqrt 3$ and $\omega_{0}^{34}=\frac{1}{\sqrt{3}}\delta_{\alpha_3\alpha_4}\,(\alpha=1,2,3)$ correspond to flavor and
color singlets, respectively. $\chi_{{1,-m}}^{34}$ is a triplet
state of spin. $\mathcal{Y}_{\ell m}(\mathbf{k})\equiv
|\mathbf{k}|^{\ell}Y_{\ell m}(\theta_{k},\phi_{k})$ is the
$\ell$th solid harmonic polynomial. $\gamma$ is a dimensionless
constant which represents the strength of the quark pair creation from the
vacuum and can be extracted by fitting the data.

For the convenience of the calculation, one usually takes the mock state to depict the meson \cite{Hayne:1981zy}
\begin{eqnarray}\label{mockmeson}
&&\left|A(n \mbox{}^{2S+1}L \,\mbox{}_{J M_{J}})
({\textbf{K}}) \right\rangle \nonumber\\&&= \sqrt{2
E}\sum_{M_{L},M_{S}} \left\langle L M_{L} S M_{S} |
J M_{J} \right\rangle \nonumber\\&&\quad\times\int \rm d
\mathbf{k}_1\rm
d\mathbf{k}_2\delta^3\left(\textbf{K}-\mathbf{k}_1-{\mathbf{k}}_2\right)\Psi_{n
L M_{L}}\left(\mathbf{k}_1,\mathbf{k}_2\right) \nonumber\\
&&\quad\times\chi^{1 2}_{S M_{S}}\varphi^{1 2}\omega^{1 2}
\left|\;q_1\left(\mathbf{k}_1\right)
\bar{q}_2\left(\mathbf{k}_2\right)\right\rangle,
\end{eqnarray}
which satisfies the normalization conditions $
\langle A(\textbf{K})|A(\textbf{K}') \rangle = 2E\,
\,\delta^3(\textbf{K}-\textbf{K}')$. Here, $\Psi_{n L M_{L}}\left(\mathbf{k}_1,\mathbf{k}_2\right)$
is the spatial wave function describing the meson.

Taking the center of the mass frame of the meson $A$: $\textbf{K}_A=0$
and $\textbf{K}_B=-\textbf{K}_C=\textbf{K}$, further we obtain a general expression of eq. (\ref{smatrix})
\begin{eqnarray}\label{T-matrix}
&&\langle BC|T|A\rangle=\sqrt{8 E_A E_B
E_C}\;\;\gamma\!\!\!\!\!\!\!\!\!\!\!
\sum_{\renewcommand{\arraystretch}{.5}\begin{array}[t]{l}
\scriptstyle M_{L_A},M_{S_A},\\
\scriptstyle M_{L_B},M_{S_B},\\
\scriptstyle M_{L_C},M_{S_C},m
\end{array}}\renewcommand{\arraystretch}{1}\!\!\!\!\!\!\!\!
\langle 1\;m;1\;-m|\;0\;0 \rangle \langle
L_A M_{L_A} S_A M_{S_A} | J_A M_{J_A}
\rangle\nonumber\\&&\quad\times\langle L_B M_{L_B} S_B M_{S_B} | J_B
M_{J_B} \rangle\langle L_C M_{L_C} S_C
M_{S_C} | J_C M_{J_C} \rangle\langle\varphi^{1 3}_B \varphi^{2 4}_C
| \varphi^{1 2}_A \varphi^{3 4}_0
 \rangle
  \nonumber\\&&\quad\times
 \langle \chi^{1 3}_{S_B M_{S_B}}\chi^{2 4}_{S_C
M_{S_C}}  | \chi^{1 2}_{S_A M_{S_A}} \chi^{3 4}_{1 -\!m} \rangle
I^{M_{L_A},m}_{M_{L_B},M_{L_C}}({\textbf{K}}) .
\end{eqnarray}
The color matrix element $ \langle\omega^{1 3}_B \omega^{2 4}_C
| \omega^{1 2}_A \omega^{3 4}_0
 \rangle=1/3$, which cancels out the factor 3 before $\gamma$ in eq. (\ref{tmatrix}).
$\langle\varphi^{1 3}_B \varphi^{2 4}_C
| \varphi^{1 2}_A \varphi^{3 4}_0
 \rangle$ and $\langle \chi^{1 3}_{S_B M_{S_B}}\chi^{2 4}_{S_C
M_{S_C}}  | \chi^{1 2}_{S_A M_{S_A}} \chi^{3 4}_{1 -\!m} \rangle$ are the flavor matrix element and the spin matrix element, respectively.
Here, the spatial integral $I^{M_{L_A},m}_{M_{L_B},M_{L_C}}(\textbf{K})$
is
\begin{eqnarray}
I^{M_{L_A},m}_{M_{L_B},M_{L_C}}(\textbf{K}) &=& \int\!\rm
d\mathbf{k}_1\rm d\mathbf{k}_2\rm d\mathbf{k}_3\rm
d\mathbf{k}_4\,\delta^3(\mathbf{k}_1+\mathbf{k}_2)\delta^3(\mathbf{k}_3+\mathbf{k}_4)\delta^2
(\textbf{K}_B-\mathbf{k}_1-\mathbf{k}_3)\delta^3(\textbf{K}_C-\mathbf{k}_2-\mathbf{k}_4)\nonumber\\
&&\times\Psi^*_{n_B L_B
M_{L_B}}(\mathbf{k}_1,\mathbf{k}_3)\Psi^*_{n_C L_C
M_{L_C}}(\mathbf{k}_2,\mathbf{k}_4)
\Psi_{n_A L_A M_{L_A}}(\mathbf{k}_1,\mathbf{k}_2)
\mathcal{Y}_{1m}\Big(\frac{\mathbf{k}_3-\mathbf{k}_4}{2}\Big),
\label{integral}
\end{eqnarray}
which describes the overlap of the
initial meson ($A$) and the created pair with the two final mesons
($B$ and $C$).

In this work, we use the simple harmonic oscillator (HO) wave function
to represent the radial portions of the meson
space wavefunction. The wave functions corresponding to the states with $nL=1S,\,1P,\,2P$ are respectively
\begin{eqnarray}
\psi_{n=1,L=0}(\mathbf{k})&=&\frac{R^{3/2}}{\pi^{3/4}}\exp\Big(-\frac{R^2\mathbf{k}^2}{2}\Big),\\
\psi_{n=1,L=1}(\mathbf{k})&=&-i2\sqrt{\frac{2}{3}}\frac{R^{5/2}}{\pi^{1/4}}\mathcal{Y}_{1}^m(\mathbf{k})
\exp\Big(-\frac{R^2\mathbf{k}^2}{2}\Big),\\
\psi_{n=2,L=1}(\mathbf{k})&=&i\frac{2R^{5/2}}{\sqrt{15}\pi^{1/4}}(5-2\mathbf{k}^2R^2)
\mathcal{Y}_{1}^m(\mathbf{k})\exp\Big(-\frac{R^2\mathbf{k}^2}{2}\Big),
\end{eqnarray}
which satisfy the normalization $\int\psi_{n,L}^*(\mathbf{k})\psi_{n,L}(\mathbf{k})d\mathbf{k}=1$. Here the solid harmonic polynomial $\mathcal{Y}_{1}^m(\mathbf{k})=\sqrt{\frac{3}{4\pi}}\,{\mbox{\boldmath $\epsilon$}}_{-m}\cdot \mathbf{k}$ with $\epsilon_{\pm1}=(\pm{1}/{\sqrt{2}},-{i}/{\sqrt{2}},0)$ and $\epsilon_0=(0,0,1)$.
$\mathbf{k}=({m_i \mathbf{k}_j-m_j\mathbf{k}_i})/({m_i+m_j})$ is the relative momentum between the quark and the antiquark within a meson when considering the quark mass difference. These HO wave functions are relevant to the calculation of the strong decay of P-wave states with the first radial excitation.

The helicity amplitude satisfies the relation
\begin{eqnarray}
\langle{}BC|T|A\rangle=\delta^3(\mathbf{K}_B+\mathbf{K}_C-\mathbf{K}_A)\mathcal{M}^{M_{J_A}M_{J_B}M_{J_C}}.
\end{eqnarray}
In terms of the partial wave amplitude, one obtains the partial decay width
\begin{eqnarray}
\Gamma = \pi^2 \frac{{|\textbf{K}|}}{M_A^2}\sum_{JL}\Big
|\mathcal{M}^{J L}\Big|^2,
\end{eqnarray}
where $|\textbf{K}|$ denotes the three momentum of
the daughter mesons in the parent's center of mass frame. The partial wave amplitude $\mathcal{M}^{J L}$ is related to
the helicity amplitude $\mathcal{M}^{M_{J_A}
M_{J_B} M_{J_C}}$ via the Jacob-Wick formula \cite{Jacob:1959at}
\begin{eqnarray}
&&{\mathcal{M}}^{J L}(A\rightarrow BC) \nonumber\\&&=
\frac{\sqrt{2 L+1}}{2 J_A +1} \!\! \sum_{M_{J_B},M_{J_C}} \langle
L 0 J M_{J_A}|J_A  M_{J_A}\rangle \nonumber\\&&\quad\times\langle
J_B M_{J_B} J_C  M_{J_C} | J M_{J_A} \rangle \mathcal{M}^{M_{J_A}
M_{J_B} M_{J_C}}({\textbf{K}}),\label{JB}\nonumber\\
\end{eqnarray}
where $\mathbf{J}=\mathbf{J}_B+\mathbf{J}_C$ and
$\mathbf{J}_{A}+\mathbf{J}_{P}=\mathbf{J}_{B}+\mathbf{J}_C+\mathbf{L}$. The calculation of transition amplitude using the $^{3}P_{0}$ model
involves two parameters: the strength of quark pair creation from
vacuum $\gamma$ and $R$ in the harmonic oscillator wavefunction.
$\gamma$ is an universal parameter which was already fixed from
other channels as indicated in Ref. \cite{Godfrey:1986wj}. The value of $R$ is chosen to reproduce the Root
Mean Square (RMS) radius obtained by solving the schr\"{o}dinger
equation with the linear potential.

\subsection{The partial wave amplitude of two-body strong decays of P-wave states with the first radial excitation}

With the preparation mentioned above, we obtain the partial wave amplitude of the strong decays of the P-wave states with the first radial excitation, which are listed in Table \ref{decay modes}. In Table \ref{amp}, the concrete expression of the partial wave amplitude is given. The details of the spatial integral $I^{M_{L_A},m}_{M_{L_B},M_{L_C}}(\textbf{K})$ are given in the appendix.

\renewcommand{\arraystretch}{1.6}

 \begin{longtable}{c|c|c}
\toprule[1pt]
State&Decay channel&Partial wave amplitude\\\midrule[1pt]

&$0^-+0^-$&$\mathcal{M}^{00}=\frac{\alpha\sqrt{2}}{3}\sqrt{E_A E_B E_C}\gamma
\big[2I_{\pm}-I_{0}\big]$\\
&$1^-+1^-$&$\mathcal{M}^{00}=\frac{\alpha\sqrt{2}}{3\sqrt{3}}\sqrt{E_A E_B E_C}\gamma
\big[I_{0}-2I_{\pm}\big]$\\

$0^+(S^\star)$&$1^+(S)+0^-$&$\mathcal{M}^{11}=\cos\theta\big[-\frac{\alpha\sqrt{2}}{3}\sqrt{E_A E_B E_C}\gamma
\big(2I^{1-1}_{00}-I^{00}_{00}\big)\big]$\\&&$\quad\quad+\sin\theta\big[-\frac{2\alpha}{3}\sqrt{E_A E_B E_C}\gamma
\big(I^{10}_{10}-I^{01}_{10}\big)\big]$\\

&$1^+(T)+0^-$&$\mathcal{M}^{11}=-\sin\theta\big[-\frac{\alpha\sqrt{2}}{3}\sqrt{E_A E_B E_C}\gamma
\big(2I^{1-1}_{00}-I^{00}_{00}\big)\big]$\\&&$\quad\quad+\cos\theta\big[-\frac{2\alpha}{3}\sqrt{E_A E_B E_C}\gamma
\big(I^{10}_{10}-I^{01}_{10}\big)\big]$\\

\midrule[1pt]

&$0^-+1^-$&$\mathcal{M}^{10}=\cos\theta'\big[\frac{\alpha}{3}\sqrt{\frac{2}{3}}\sqrt{E_A E_B E_C}\gamma
\big(2I_{\pm}-I_{0}\big)\big]$\\&&$\quad\quad\quad\quad+\sin\theta'\big[-\frac{2\alpha}{3\sqrt{3}}\sqrt{E_A E_B E_C}\gamma
\big(2I_{\pm}-I_{0}\big)\big]$\\
&&$\mathcal{M}^{12}=\cos\theta'\big[\frac{\alpha}{3\sqrt{3}}\sqrt{E_A E_B E_C}\gamma
\big(2I_{\pm}+2I_{0}\big)\big]$
\\&&$\quad\quad+\sin\theta'\big[\frac{2\alpha}{3\sqrt{6}}\sqrt{E_A E_B E_C}\gamma
\big(I_{\pm}+I_{0}\big)\big]$\\

&$1^-+0^-$&$\mathcal{M}^{10}=\cos\theta'\big[\frac{\alpha}{3}\sqrt{\frac{2}{3}}\sqrt{E_A E_B E_C}\gamma
\big(2I_{\pm}-I_{0}\big)\big]$\\&&$\quad\quad\quad\quad+\sin\theta'\big[-\frac{2\alpha}{3\sqrt{3}}\sqrt{E_A E_B E_C}\gamma
\big(2I_{\pm}-I_{0}\big)\big]$\\
$1^+(S^\star)$&&$\mathcal{M}^{12}=\cos\theta'\big[\frac{\alpha}{3\sqrt{3}}\sqrt{E_A E_B E_C}\gamma
\big(2I_{\pm}+2I_{0}\big)\big]$
\\&&$\quad\quad+\sin\theta'\big[\frac{2\alpha}{3\sqrt{6}}\sqrt{E_A E_B E_C}\gamma
\big(I_{\pm}+I_{0}\big)\big]$\\


&$1^-+1^-$&$\mathcal{M}^{10}=\cos\theta'\big[-\frac{2\alpha}{3}\sqrt{\frac{1}{3}}\sqrt{E_A E_B E_C}\gamma
\big(2I_{\pm}-I_{0}\big)\big]$\\
&&$\mathcal{M}^{22}=\sin\theta'\big[\frac{2\alpha}{3}\sqrt{E_A E_B E_C}\gamma\big(I_\pm+I_{0}\big)\big]$\\

&$0^++0^-$&$\mathcal{M}^{01}=\cos\theta'\big[\frac{\alpha}{3}\sqrt{\frac{2}{3}}\sqrt{E_A E_B E_C}\gamma
\big (I^{00}_{00}+2I^{01}_{10}\big)\big]$\\
&&$\quad\quad+\sin\theta'\big[\frac{2\alpha}{3\sqrt{3}}\sqrt{E_A E_B E_C}\gamma
\big (I^{1-1}_{00}+I^{10}_{10}\big)\big]$\\

&$1^+(S)+0^-$&$\mathcal{M}^{11}=\sin\theta\cos\theta'\big[\frac{\alpha}{3\sqrt{2}}\sqrt{E_A E_B E_C}\gamma
\big(2I^{1-1}_{00}+2I^{10}_{10}\big)\big]$\\
&&$\quad\quad+\cos\theta\sin\theta'\big[-\frac{\alpha\sqrt{2}}{3}\sqrt{E_A E_B E_C}\gamma
\big (I^{10}_{10}-I^{01}_{10}\big)\big]$\\
&&$\quad\quad\quad\quad\quad\,\,+\sin\theta\sin\theta'\big[-\frac{\alpha}{3}\sqrt{E_A E_B E_C}\gamma
\big(-I^{00}_{00}-I^{01}_{10}+I^{1-1}_{00}\big)\big]$\\

&$1^+(T)+0^-$&$\mathcal{M}^{11}=\cos\theta\cos\theta'\big[\frac{\alpha}{3\sqrt{2}}\sqrt{E_A E_B E_C}\gamma
\big(2I^{1-1}_{00}+2I^{10}_{10}\big)\big]$\\
&&$\quad\quad-\sin\theta\sin\theta'\big[-\frac{\alpha\sqrt{2}}{3}\sqrt{E_A E_B E_C}\gamma
\big (I^{10}_{10}-I^{01}_{10}\big)\big]$\\
&&$\quad\quad\quad\quad\quad\,\,+\cos\theta\sin\theta'\big[-\frac{\alpha}{3}\sqrt{E_A E_B E_C}\gamma
\big(-I^{00}_{00}-I^{01}_{10}+I^{1-1}_{00}\big)\big]$\\
&$2^++0^-$&$\mathcal{M}^{21}=\cos\theta'\big[\frac{\alpha}{3\sqrt{30}}\sqrt{E_A E_B E_C}\gamma\big(
4I^{00}_{00}-4I^{01}_{10}-6I^{1-1}_{00}
+6I^{10}_{10}\big)\big]$\\
&&$\quad\quad+\sin\theta'\big[-\frac{\alpha}{3\sqrt{15}}\sqrt{E_A E_B E_C}\gamma\big(
3I^{00}_{00}-3I^{01}_{10}-7I^{1-1}_{00}
+2I^{10}_{10}\big)\big]$\\\midrule[1pt]

&$0^-+1^-$&$\mathcal{M}^{10}=-\sin\theta'\big[\frac{\alpha}{3}\sqrt{\frac{2}{3}}\sqrt{E_A E_B E_C}\gamma
\big(2I_{\pm}-I_{0}\big)\big]$\\&&$\quad\quad\quad\quad+\cos\theta'\big[-\frac{2\alpha}{3\sqrt{3}}\sqrt{E_A E_B E_C}\gamma
\big(2I_{\pm}-I_{0}\big)\big]$\\
&&$\mathcal{M}^{12}=-\sin\theta'\big[\frac{\alpha}{3\sqrt{3}}\sqrt{E_A E_B E_C}\gamma
\big(2I_{\pm}+2I_{0}\big)\big]$
\\&&$\quad\quad+\cos\theta'\big[\frac{2\alpha}{3\sqrt{6}}\sqrt{E_A E_B E_C}\gamma
\big(I_{\pm}+I_{0}\big)\big]$\\

&$1^-+0^-$&$\mathcal{M}^{10}=-\sin\theta'\big[\frac{\alpha}{3}\sqrt{\frac{2}{3}}\sqrt{E_A E_B E_C}\gamma
\big(2I_{\pm}-I_{0}\big)\big]$\\&&$\quad\quad\quad\quad+\cos\theta'\big[-\frac{2\alpha}{3\sqrt{3}}\sqrt{E_A E_B E_C}\gamma
\big(2I_{\pm}-I_{0}\big)\big]$\\
$1^+(T^\star)$&&$\mathcal{M}^{12}=-\sin\theta'\big[\frac{\alpha}{3\sqrt{3}}\sqrt{E_A E_B E_C}\gamma
\big(2I_{\pm}+2I_{0}\big)\big]$
\\&&$\quad\quad+\cos\theta'\big[\frac{2\alpha}{3\sqrt{6}}\sqrt{E_A E_B E_C}\gamma
\big(I_{\pm}+I_{0}\big)\big]$\\


&$1^-+1^-$&$\mathcal{M}^{10}=-\sin\theta'\big[-\frac{2\alpha}{3}\sqrt{\frac{1}{3}}\sqrt{E_A E_B E_C}\gamma
\big(2I_{\pm}-I_{0}\big)\big]$\\
&&$\mathcal{M}^{22}=\cos\theta'\big[\frac{2\alpha}{3}\sqrt{E_A E_B E_C}\gamma\big(I_\pm+I_{0}\big)\big]$\\

&$0^++0^-$&$\mathcal{M}^{01}=-\sin\theta'\big[\frac{\alpha}{3}\sqrt{\frac{2}{3}}\sqrt{E_A E_B E_C}\gamma
\big (I^{00}_{00}+2I^{01}_{10}\big)\big]$\\
&&$\quad\quad+\cos\theta'\big[\frac{2\alpha}{3\sqrt{3}}\sqrt{E_A E_B E_C}\gamma
\big (I^{1-1}_{00}+I^{10}_{10}\big)\big]$\\

&$1^+(S)+0^-$&$\mathcal{M}^{11}=-\sin\theta\sin\theta'\big[\frac{\alpha}{3\sqrt{2}}\sqrt{E_A E_B E_C}\gamma
\big(2I^{1-1}_{00}+2I^{10}_{10}\big)\big]$\\
&&$\quad\quad+\cos\theta\cos\theta'\big[-\frac{\alpha\sqrt{2}}{3}\sqrt{E_A E_B E_C}\gamma
\big (I^{10}_{10}-I^{01}_{10}\big)\big]$\\
&&$\quad\quad\quad\quad\quad\,\,+\sin\theta\cos\theta'\big[-\frac{\alpha}{3}\sqrt{E_A E_B E_C}\gamma
\big(-I^{00}_{00}-I^{01}_{10}+I^{1-1}_{00}\big)\big]$\\

&$1^+(T)+0^-$&$\mathcal{M}^{11}=-\cos\theta\sin\theta'\big[\frac{\alpha}{3\sqrt{2}}\sqrt{E_A E_B E_C}\gamma
\big(2I^{1-1}_{00}+2I^{10}_{10}\big)\big]$\\
&&$\quad\quad-\sin\theta\cos\theta'\big[-\frac{\alpha\sqrt{2}}{3}\sqrt{E_A E_B E_C}\gamma
\big (I^{10}_{10}-I^{01}_{10}\big)\big]$\\
&&$\quad\quad\quad\quad\quad\,\,+\cos\theta\cos\theta'\big[-\frac{\alpha}{3}\sqrt{E_A E_B E_C}\gamma
\big(-I^{00}_{00}-I^{01}_{10}+I^{1-1}_{00}\big)\big]$\\
&$2^++0^-$&$\mathcal{M}^{21}=-\sin\theta'\big[\frac{\alpha}{3\sqrt{30}}\sqrt{E_A E_B E_C}\gamma\big(
4I^{00}_{00}-4I^{01}_{10}-6I^{1-1}_{00}
+6I^{10}_{10}\big)\big]$\\
&&$\quad\quad+\cos\theta'\big[-\frac{\alpha}{3\sqrt{15}}\sqrt{E_A E_B E_C}\gamma\big(
3I^{00}_{00}-3I^{01}_{10}-7I^{1-1}_{00}
+2I^{10}_{10}\big)\big]$\\\midrule[1pt]

&$0^-+0^-$&$\mathcal{M}^{02}=\frac{2\alpha}{3\sqrt{5}}\sqrt{E_A E_B E_C}\gamma\big[
I_{\pm}+I_{0}\big]$\\

&$0^-+1^-$&$\mathcal{M}^{12}=\frac{2\alpha}{\sqrt{30}}\sqrt{E_A E_B E_C}\gamma\big[
I_{\pm}+I_{0}\big]$\\

&$1^-+0^-$&$\mathcal{M}^{12}=\frac{2\alpha}{\sqrt{30}}\sqrt{E_A E_B E_C}\gamma\big[
I_{\pm}+I_{0}\big]$\\

$2^+(T^\star)$&$1^-+1^-$&$\mathcal{M}^{20}=\frac{2\alpha}{3}\sqrt{\frac{2}{3}}\sqrt{E_A E_B E_C}\gamma\big[
2I_{\pm}-I_{0}\big]$\\

&$1^+(S)+0^-$&$\mathcal{M}^{11}=\cos\theta\big[\frac{\alpha}{15\sqrt{2}}\sqrt{E_A E_B E_C}\gamma
\big (4I^{00}_{00}+6I^{01}_{10}+4I^{1-1}_{00}+6I^{10}_{10}\big)\big]$\\
&&$\quad\quad+\sin\theta\big[\frac{\alpha}{15}\sqrt{E_A E_B E_C}\gamma
\big (3I^{00}_{00}+7I^{01}_{10}+3I^{1-1}_{00}+2I^{10}_{10}\big)\big]$\\
&$1^+(T)+0^-$&$\mathcal{M}^{11}=-\sin\theta\big[\frac{\alpha}{15\sqrt{2}}\sqrt{E_A E_B E_C}\gamma
\big (4I^{00}_{00}+6I^{01}_{10}+4I^{1-1}_{00}+6I^{10}_{10}\big)\big]$\\
&&$\quad\quad+\cos\theta\big[\frac{\alpha}{15}\sqrt{E_A E_B E_C}\gamma
\big (3I^{00}_{00}+7I^{01}_{10}+3I^{1-1}_{00}+2I^{10}_{10}\big)\big]$\\


&$2^++0^-$&$\mathcal{M}^{21}=\frac{\alpha}{5\sqrt{3}}\sqrt{E_A E_B E_C}\gamma
\big [I^{00}_{00}-I^{01}_{10}+I^{1-1}_{00}+4I^{10}_{10}\big]$\\

\bottomrule[1pt]
\caption{The expression of the partial wave amplitude for the strong decays of
P-wave states with the first radial excitation. Here $\alpha=2/\sqrt{18},\,-1/3$ are for the strong decay involved in $\eta$ and $\eta^\prime$ mesons respectively while
$\alpha=1/\sqrt{3}$ is for the other strong decays, which are the result from the flavor matrix element.
\label{amp}}
\end{longtable}

\subsection{Numerical result}

The input parameters involved in the $^3P_0$ model include the strength of quark pair creation from the vacuum, the $R$ value in the HO wave function and the mass of the meson.
One takes the strength of quark pair creation from the vacuum as $\gamma=6.3$ \cite{Godfrey:1986wj}, which is $\sqrt{96\pi}$ times larger than that adopted by the other theoretical groups
\cite{Close:2005se,Kokoski:1985is}. The strength of $s\bar{s}$ creation satisfies
$\gamma_{s}=\gamma/\sqrt{3}$ \cite{LeYaouanc:1977gm}. If reproducing the realistic root mean square (RMS)
radius by solving the schr\"{o}dinger equation with the linear
potential, one can obtain the value of $R$ in the HO wave function \cite{Godfrey:1986wj}. The mass and the $R$ value used in this work are shown in Table \ref{parameter}.

\renewcommand{\arraystretch}{1.2}

\begin{table}[htb]
\begin{tabular}{c|cccccc}\toprule[1pt]
State& mass (MeV) \cite{Amsler:2008zzb}& R (GeV$^{-1}$) \cite{Godfrey:1986wj}\\\midrule[1pt]
$D$& 1869.62 ($\pm$)\; 1864.84($0$)&1.52\\
$D_s$&1968.49 ($\pm$)&1.41\\
$D^*$& 2010.27 ($\pm$)\; 2006.97($0$)&1.85\\
$D_s^*$&2112.3 ($\pm$)&1.69\\
$D_0^*(2400)$& 2403 $(\pm)\;$2352 $(0)$ &1.85\\
$D_{s0}^*(2317)$& 2317.8 $(\pm)$ &1.75\\
$D_{s1}(2460)$& 2459.6 $(\pm)$ &1.92\\
$D_1(2430)$& 2427 $(\pm)\;$2427 $(0)$ &2.00\\
$D_1(2420)$& 2423.4 $(\pm)\;$2422.3 $(0)$ &2.00\\
$D_2^*(2460)$& 2460.1 $(\pm)\;$2461.1 $(0)$ &2.22\\
$K$& 493.677$(\pm)$\;497.614 $(0)$&1.41\\
$K^*$& 891.66$(\pm)$\;896.00 $(0)$&2.08\\
$\eta$&547.853&1.41\\
$\eta'$&957.66&1.41\\
$\phi$&1019.455&2.08\\
\bottomrule[1pt]
\end{tabular}\caption{The $R$ value in the HO wave function and the mass relevant to the strong decays listed in Table \ref{decay modes}. Here $(\pm)$ and $(0)$ denote the charge of the meson.
\label{parameter}}
\end{table}

\begin{center}
\begin{figure}[htb]
\begin{tabular}{ccc}
\scalebox{0.8}{\includegraphics{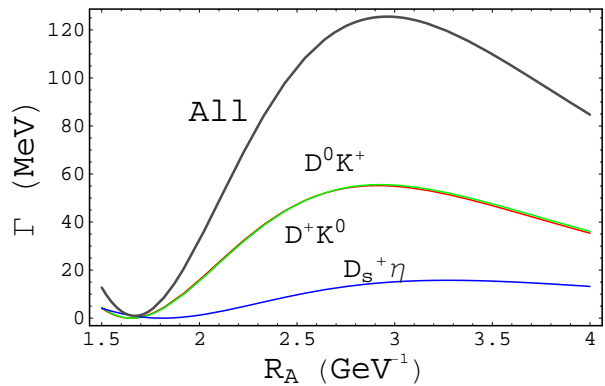}}
\end{tabular}
\caption{The variation of the strong decay mode $0^+(S^{\star})\to 0^-+0^-$ with $R_{A}$.
\label{aa}}
\end{figure}
\end{center}

If the mass of the charmed-strange meson with $0^+(S^\star)$ is 2.837 GeV predicted in Ref. \cite{Matsuki:2006rz}, there only exists the decay channel $0^+(S^\star)\to 0^-+0^-$, which is allowed by the phase space. In Fig. \ref{aa}, we give the dependence of the partial decay widths of the strong decay of $0^+(S^\star)$ state on the $R_A$. Here, $R_A$ is the $R$ value of the HO wave function of charmed-strange state with $0^+(S^\star)$. The minimum of the decay width around $R_A=1.7$ GeV$^{-1}$ in Fig. \ref{aa} is due to the node in the radial wave function of $0^+(S^\star)$.
The total decay width of $0^+(S^\star)$ charmed-strange meson is 108 MeV with $R_A=2.8$ GeV$^{-1}$.

\begin{center}
\begin{figure}[htb]
\begin{tabular}{ccc}
\scalebox{0.56}{\includegraphics{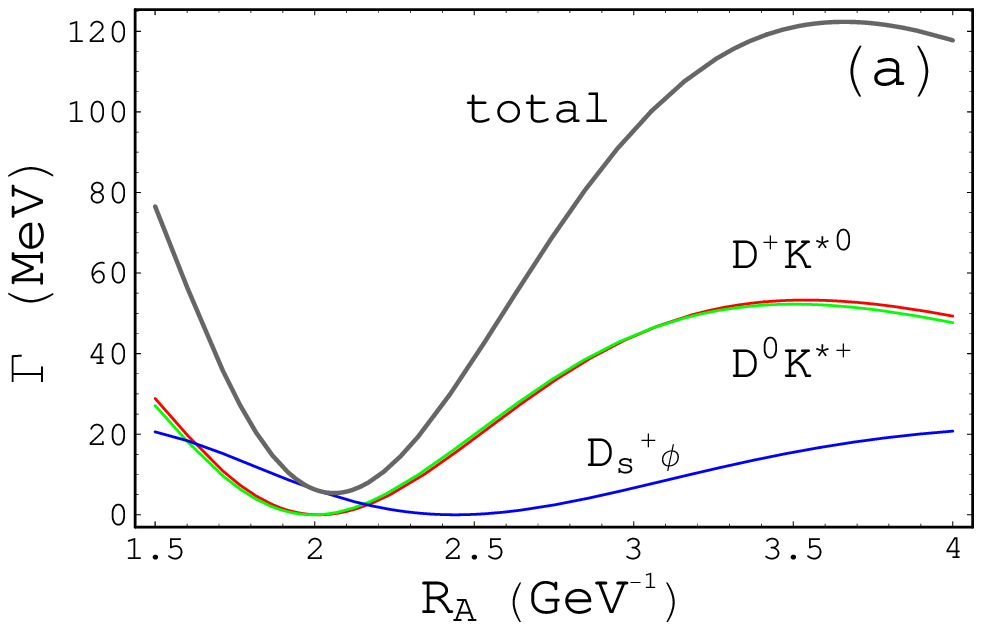}}&
\scalebox{0.56}{\includegraphics{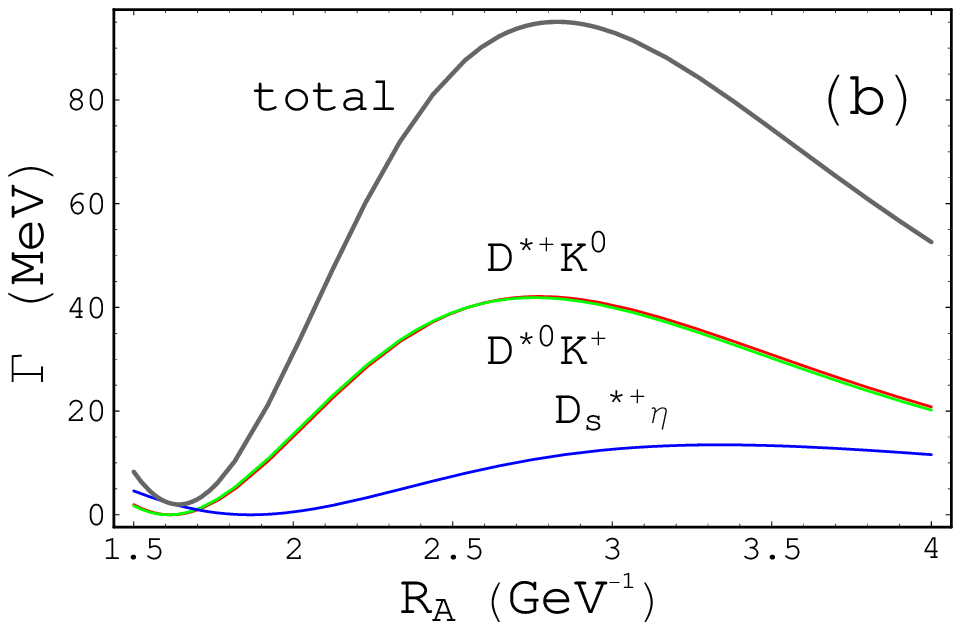}}&
\scalebox{0.56}{\includegraphics{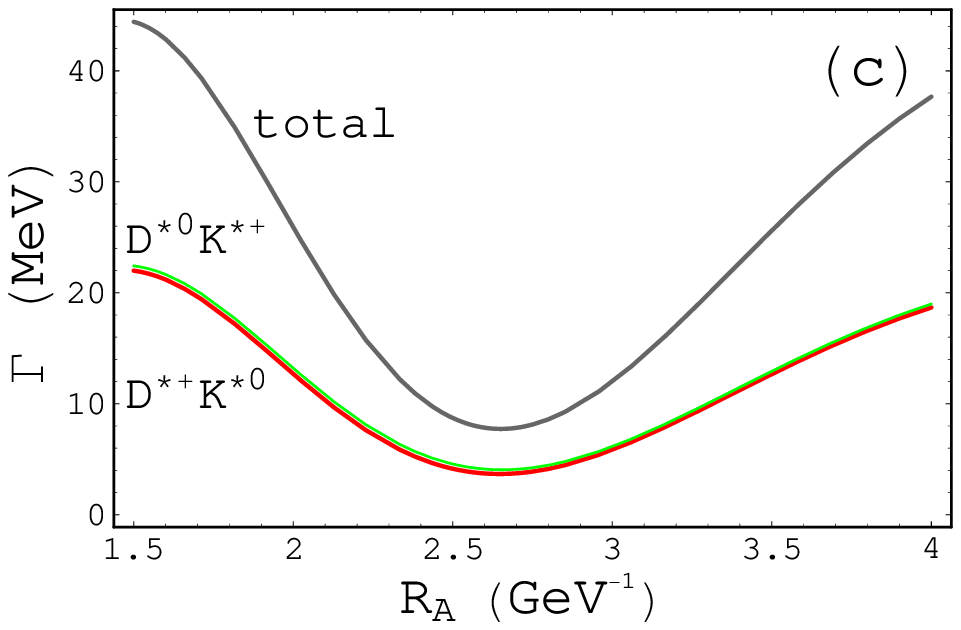}}\\
\scalebox{0.56}{\includegraphics{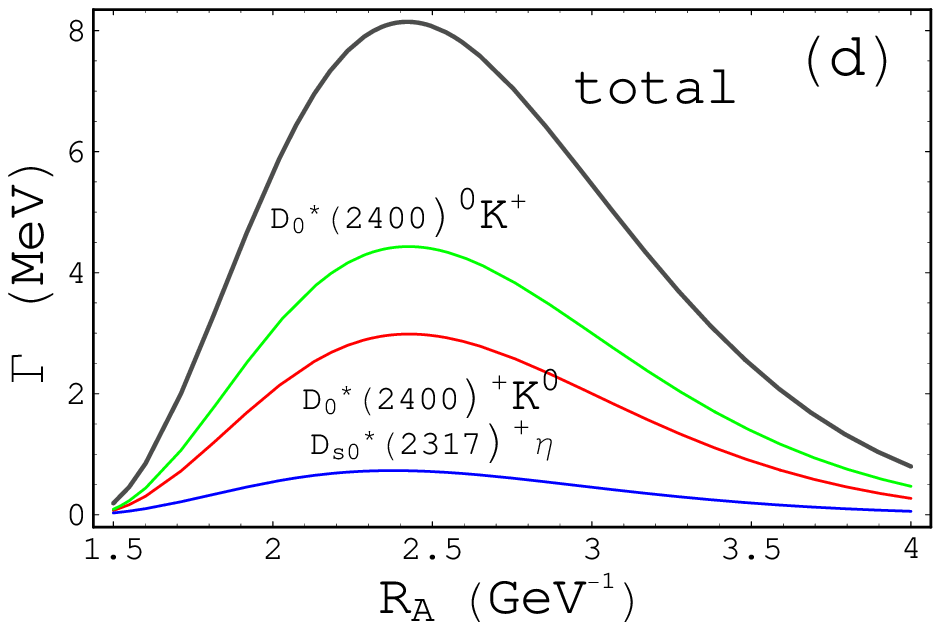}}&
\scalebox{0.56}{\includegraphics{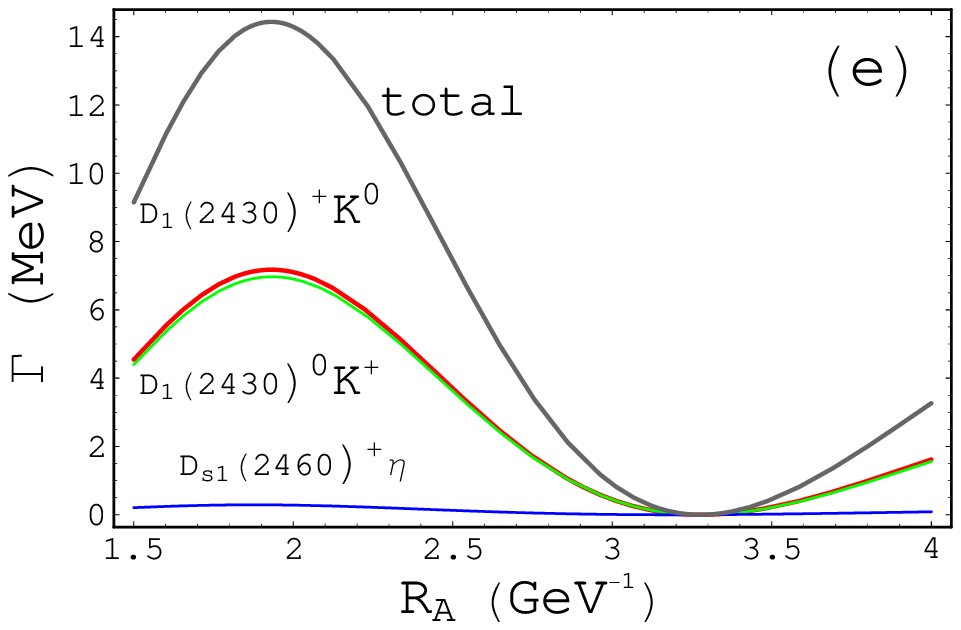}}&\scalebox{0.56}{\includegraphics{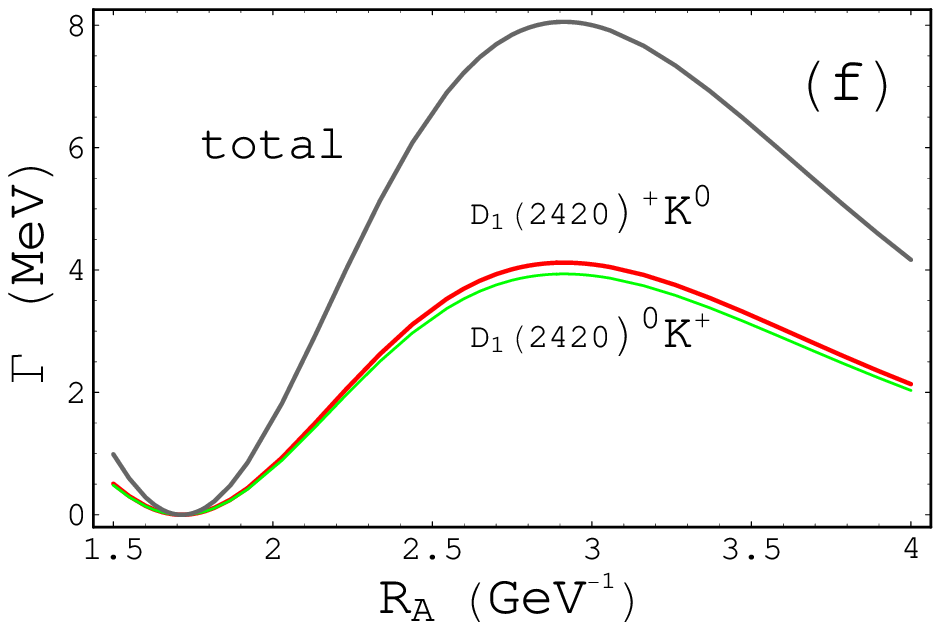}}\\
\scalebox{0.56}{\includegraphics{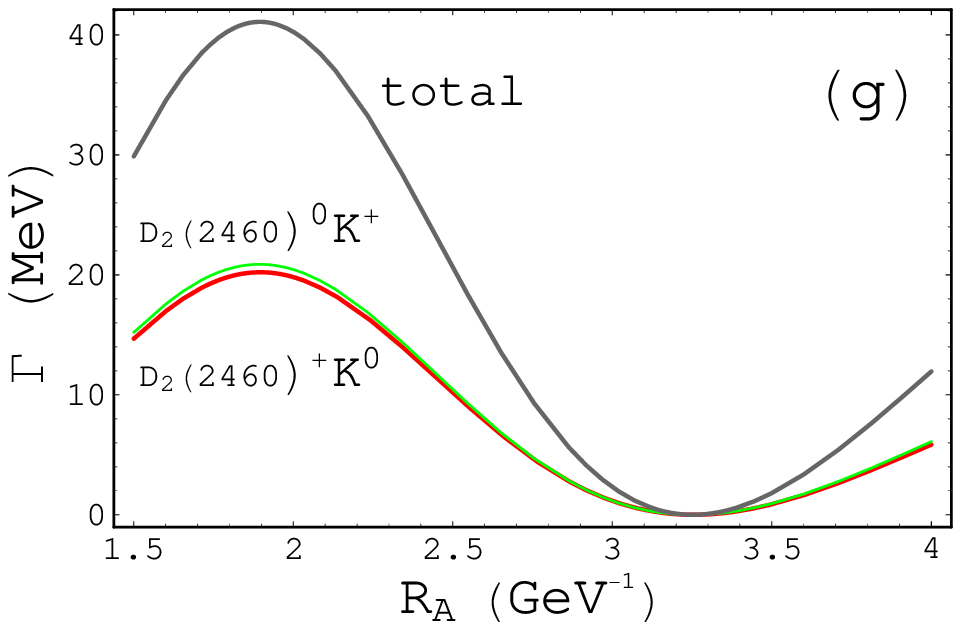}}&
\end{tabular}
\caption{The variation of the strong decays for (a)
$1^+(S^\star)\to 0^-+1^-$, (b) $1^+(S^\star)\to 1^-+0^-$, (c) $1^+(S^\star)\to 1^-+1^-$, (d) $1^+(S^\star)\to 0^++0^-$,
(e) $1^+(S^\star)\to 1^+(S)+0^-$, (f) $1^+(S^\star)\to 1^+(T)+0^-$ and (g) $1^+(S^\star)\to 2^++0^-$ with the factor $R_A$ of the HO wavefunction of
$1^+(S^\star)$. Here the total partial decay width is labeled by "total" in diagrams. \label{1s}}
\end{figure}
\end{center}
\begin{center}
\begin{figure}[htb]
\begin{tabular}{ccc}
\scalebox{0.56}{\includegraphics{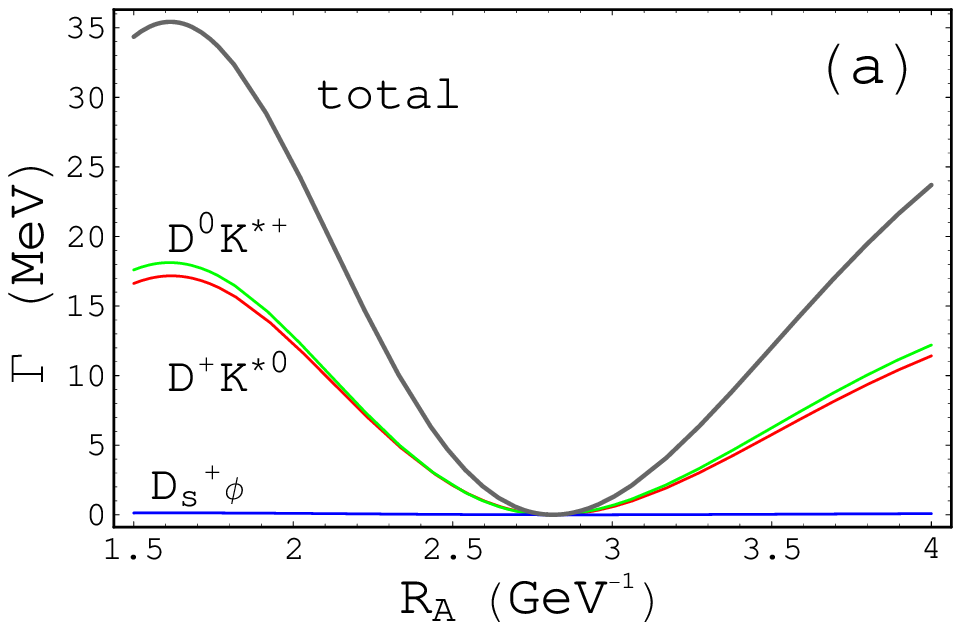}}&
\scalebox{0.56}{\includegraphics{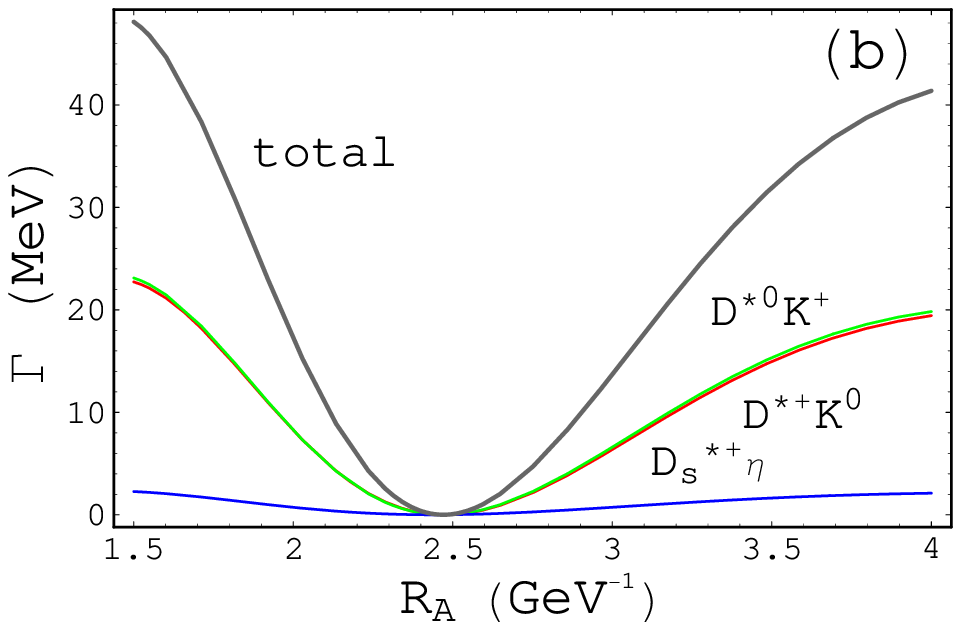}}&
\scalebox{0.56}{\includegraphics{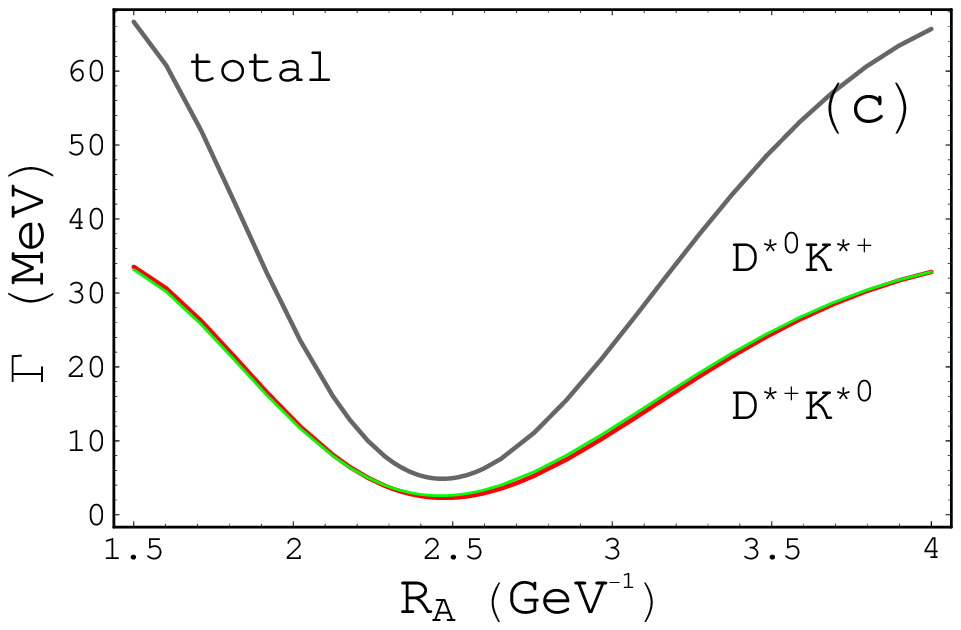}}\\
\scalebox{0.56}{\includegraphics{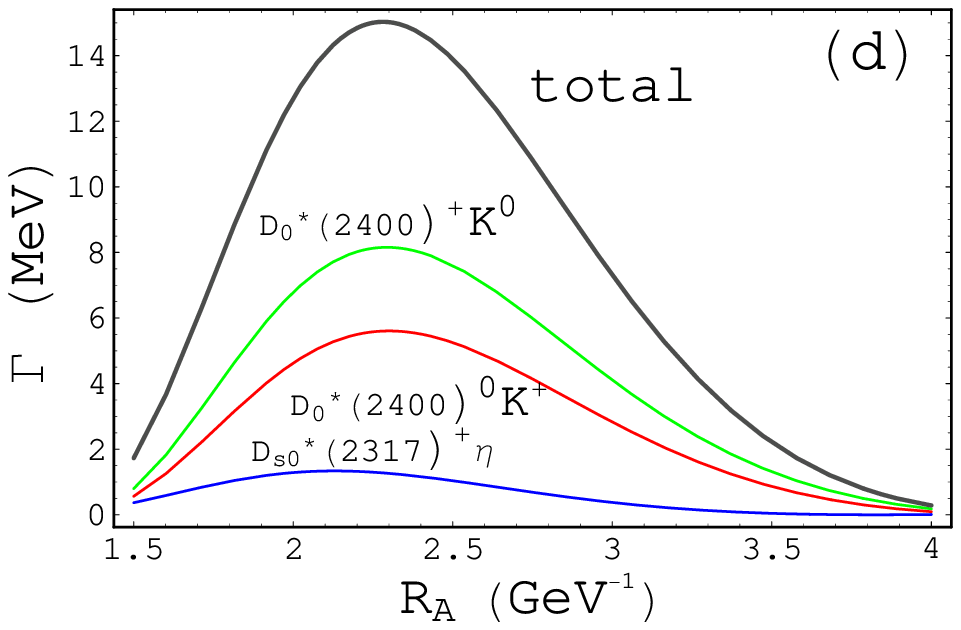}}&
\scalebox{0.56}{\includegraphics{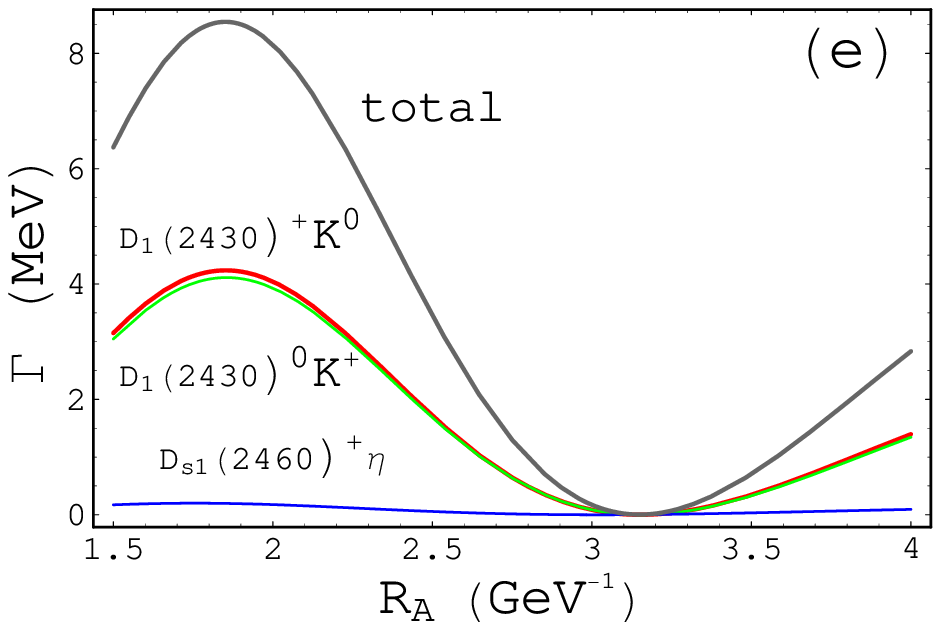}}&\scalebox{0.56}{\includegraphics{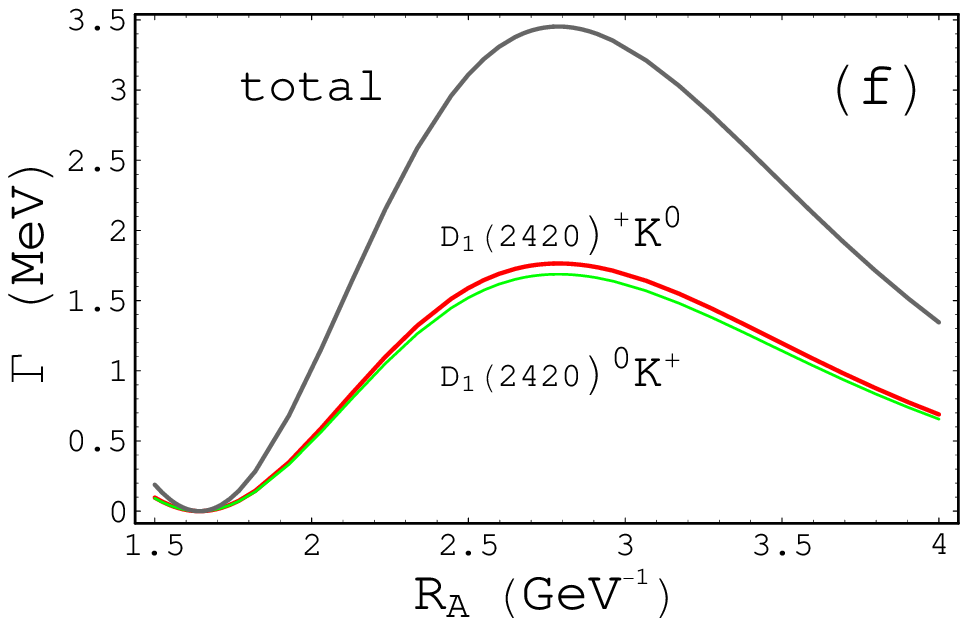}}\\
\scalebox{0.56}{\includegraphics{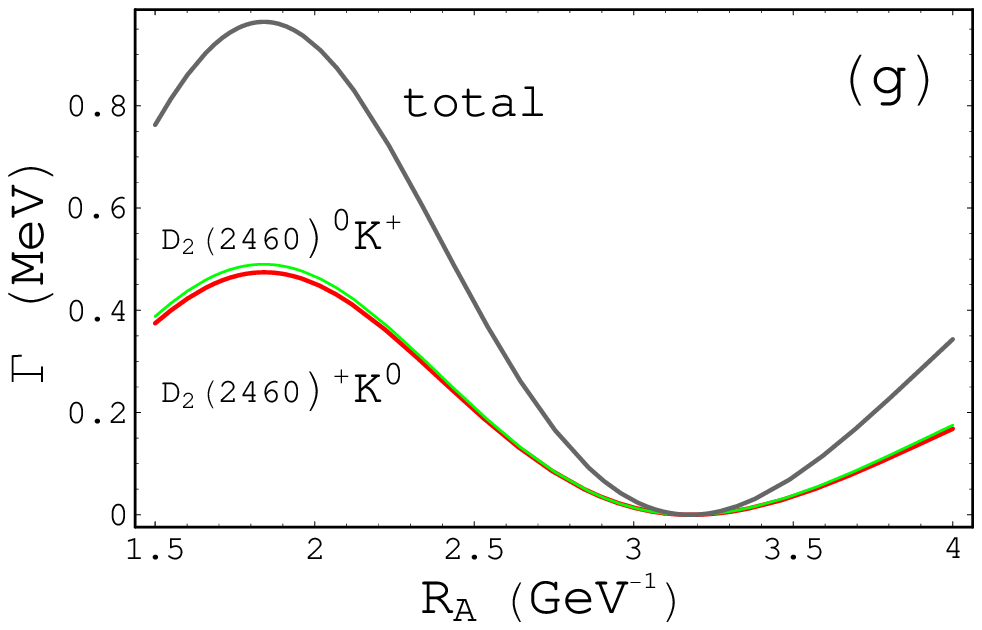}}
\end{tabular}
\caption{The variation of the strong decays for (a)
$1^+(T^\star)\to 0^-+1^-$, (b) $1^+(T^\star)\to 1^-+0^-$, (c) $1^+(T^\star)\to 1^-+1^-$, (d) $1^+(T^\star)\to 0^++0^-$,
(e) $1^+(T^\star)\to 1^+(S)+0^-$, (f) $1^+(T^\star)\to 1^+(T)+0^-$ and (g) $1^+(T^\star)\to 2^++0^-$ with the factor $R_A$ of the HO wavefunction of
$1^+(T^\star)$. Here the total partial decay width is labeled by "total" in diagrams. \label{1t}}
\end{figure}
\end{center}
\begin{center}
\begin{figure}[htb]
\begin{tabular}{ccc}
\scalebox{1.2}{\includegraphics{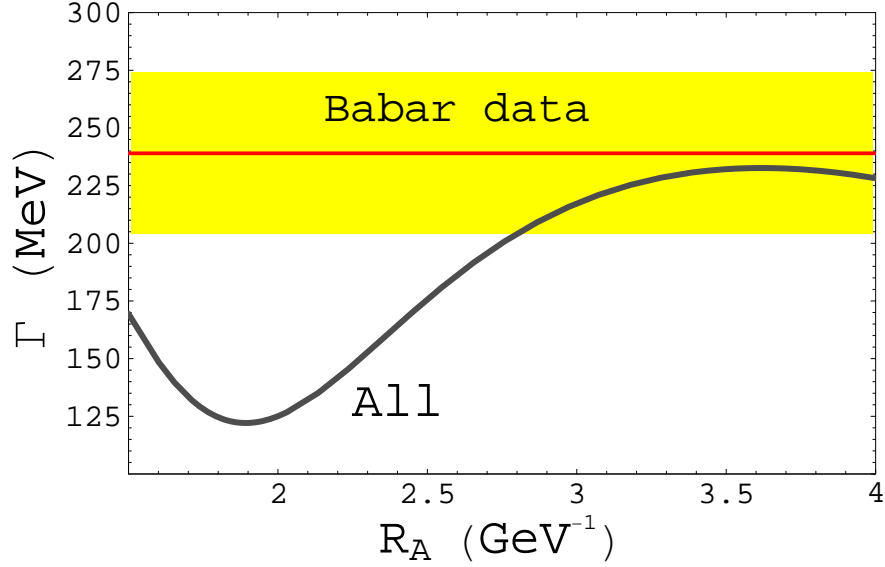}}
\end{tabular}
\caption{
A comparison of the total decay width of $1^+(S^\star)$ with Babar data. Here the red line and the yellow band are the central value the error of the total width of $D_{sJ}(3040)$ measured by Babar. \label{compare-1s}}
\end{figure}
\end{center}

In this work, we take the masses of $1^+(S^\star)$ and $1^+(T^\star)$ charmed-strange mesons as $3.044$ GeV, which is the experimental value of the mass of $D_{sJ}(3040)^+$. Then, we calculate the decay of $D_{sJ}(3040)^+$ under the two assumptions $1^+(S^\star)$ and $1^+(T^\star)$. In Figs. \ref{1s} and \ref{1t}, we present the numerical results of the two charmed-strange mesons $1^+(S^\star)$ and $1^+(T^\star)$.
The dependence of the total decay width of $1^{+}(S)$ on the $R_A$ is shown in Fig. \ref{compare-1s}. Here, $R_A$ denotes the $R$ value in the HO wave function of $D_{sJ}(3040)^+$.  By comparing the calculated total decay width of $D_{sJ}(3040)^+$ with that of the Babar data, one finds that the total decay width ($\sim 204$ MeV) of $D_{sJ}(3040)^+$ obtained by the $^3P_0$ model reaches up to the lower limit of the experimental width of $D_{sJ}(3040)^+$ when taking $R_A$ as 2.8 GeV$^{-1}$. With increasing the $R_A$ up to 3.5 GeV$^{-1}$, the total decay width is close to the central value of the experimental width of $D_{sJ}(3040)^+$. Thus, studying the decay width of $D_{sJ}(3040)^+$ under $1^+(S^{\star})$ assignment shows that the first radial excitation of P-wave charmed-strange meson to $D_{sJ}(3040)^+$, i.e. $1^+(S^{\star})$, is suitable.

The result of the partial decay widths of $1^+(S^\star)$ charmed-strange meson corresponding to $R_{A}=2.8$ GeV$^{-1}$ (see Fig. \ref{1s}) indicates that $0^-+1^-$ ($D^+K^{*0}$, $D^0 K^{*+}$ and $D_s^+\phi$) and $1^-+0^-$ ($D^{*+}K^0$, $D^{*0} K^+$ and $D_s^{*+}\eta$) are the dominant decay modes of $D_{sJ}(3040)^+$, which further explain why $D_{sJ}(3040)^+$ was firstly observed in $D^*K$ decay channel. An experimental search of $D_{sJ}(3040)^+$ in $0^-+1^-$ channel ($D^+K^{*0}$, $D^0 K^{*+}$ and $D_s^+\phi$) is encouraged in terms of the ratio $$\frac{\Gamma(1^+(S^\star)\to 0^-+1^-)}{
\Gamma(1^+(S^\star)\to 1^-+0^-)}\sim 0.79$$ corresponding to $R_{A}=2.8$ GeV$^{-1}$.

\begin{center}
\begin{figure}[htb]
\begin{tabular}{ccc}
\scalebox{0.85}{\includegraphics{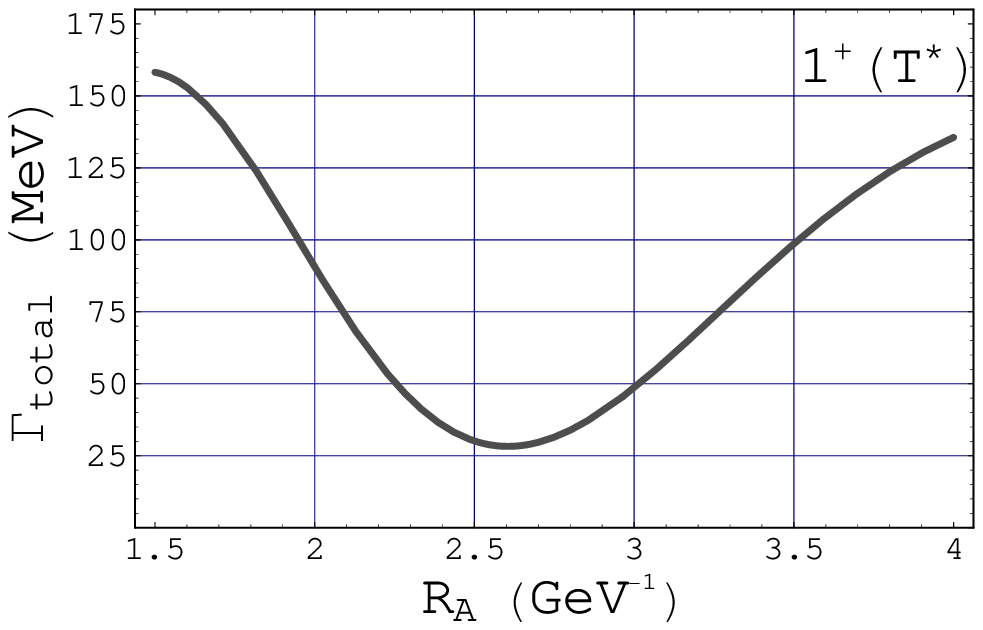}}&\scalebox{0.85}{\includegraphics{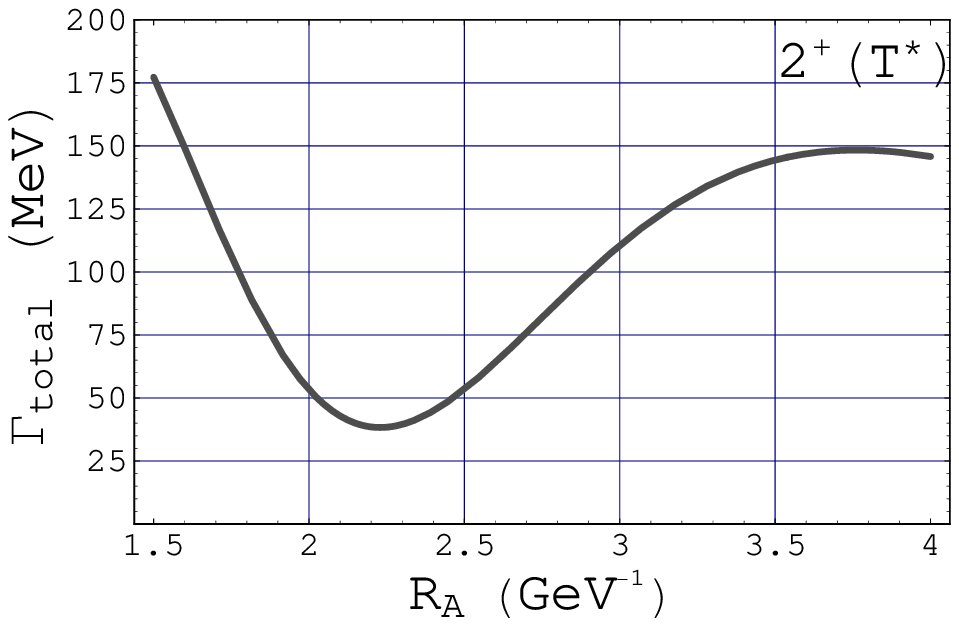}}
\end{tabular}
\caption{
The dependence of the total decay width of $1^{+}(T^\star)$ and $2^{+}(T^\star)$ states on $R_A$.
\label{compare-1t}}
\end{figure}
\end{center}

Under the assignment of $1^{+}(T^\star)$ to $D_{sJ}(3040)$, we can obtain the variation of the strong decays for
$1^+(T^\star)\to 0^-+1^-,\,1^-+0^-,\, 1^-+1^-,\, 0^++0^-,\, 1^+(S)+0^-,\, 1^+(T)+0^-,\,2^++0^-$ with the factor $R_A$ ($R$ value of the HO wavefunction of
$1^+(T^\star)$) which is depicted in Fig. \ref{1t}. Furthermore, the dependence of the total decay width on the $R_A$ value is listed in Fig. \ref{compare-1t}. The total decay width of $D_{sJ}(3040)^+$ is about 33.8 MeV with $R_A=2.8$ GeV$^{-1}$, which is consistent with our knowledge, i.e., the $1^+$ state existing $T$ doublet is of narrow width. In fact, the result of the decay of $1^{+}(T^\star)$ state further indicates that $D_{sJ}(3040)^+$ can not be explained as $1^{+}(T^\star)$ charmed-strange meson.

One also predicts that the partial decay widths corresponding to the decay channels
$1^+(T^\star)\to 0^-+1^-,\,1^-+0^-,\,1^-+1^-,\, 0^++0^-,\,1^+(S)+0^-,\,1^{+}(T)+0^-,\,2^++0^-$ are $9.8\times10^{-3}$ MeV, 6.3 MeV, 13.0 MeV, 10.1 MeV, $9.9\times10^{-1}$ MeV, 3.5 MeV and $1.3\times10^{-1}$ MeV, respectively. These numerical results show that $1^-+0^-,\,1^-+1^-,\, 0^++0^-,\,1^{+}(T)+0^-$ channels are important when searching $1^+(T^\star)$ state in experiment.

\begin{center}
\begin{figure}[htb]
\begin{tabular}{ccc}
\scalebox{0.56}{\includegraphics{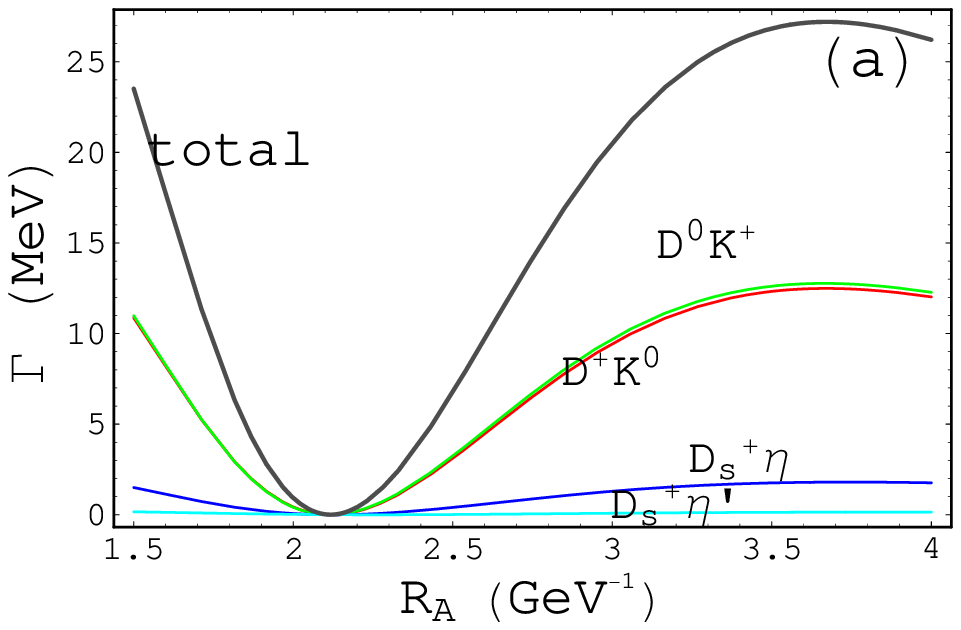}}&
\scalebox{0.56}{\includegraphics{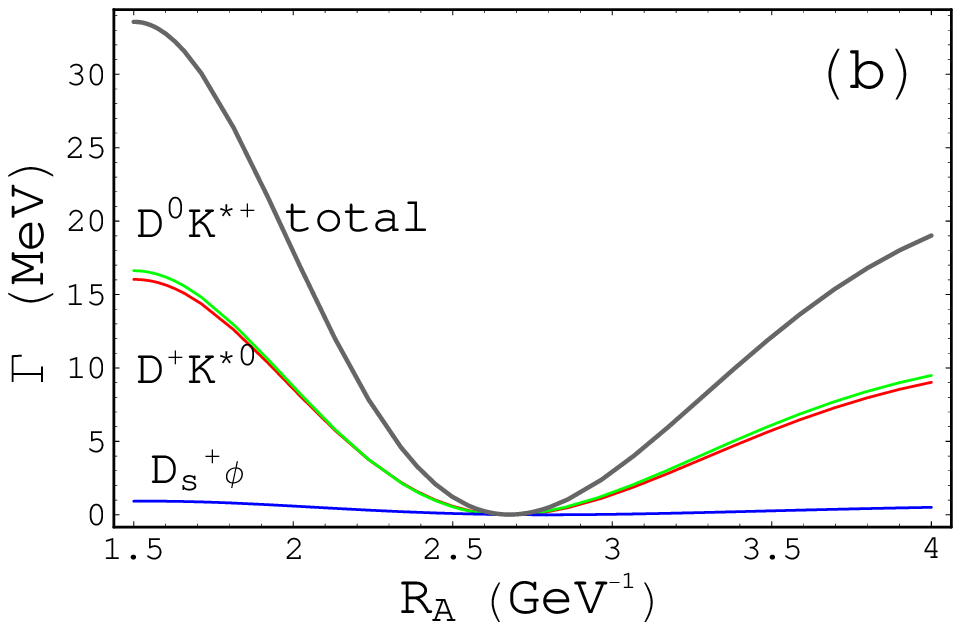}}&
\scalebox{0.56}{\includegraphics{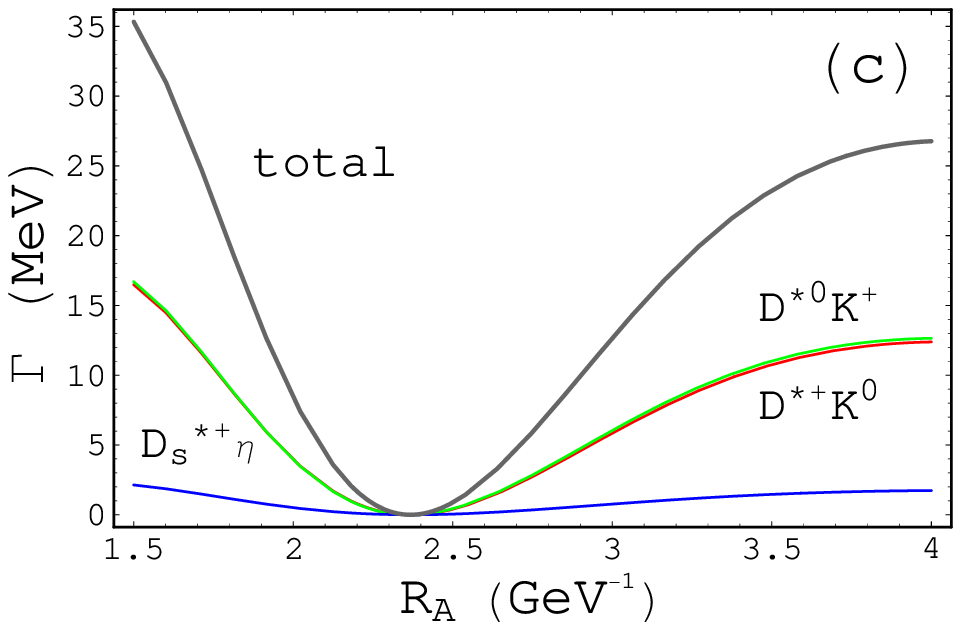}}\\
\scalebox{0.56}{\includegraphics{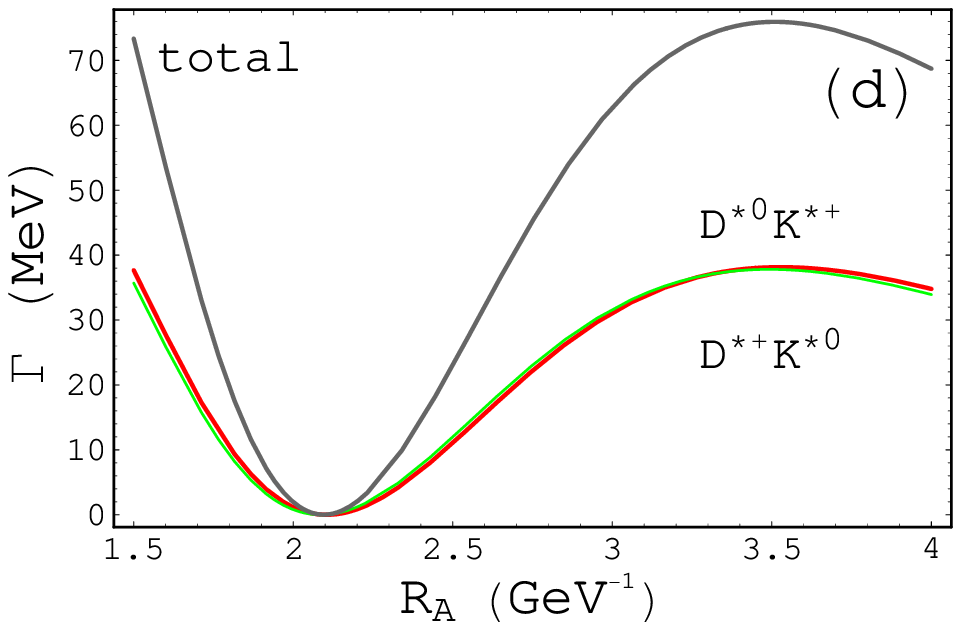}}&
\scalebox{0.56}{\includegraphics{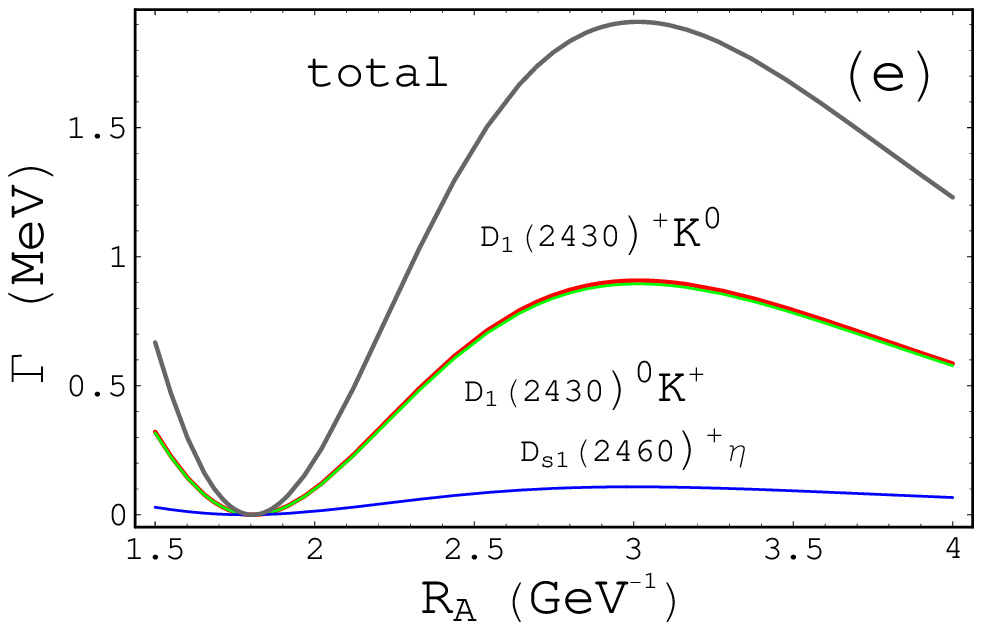}}&\scalebox{0.56}{\includegraphics{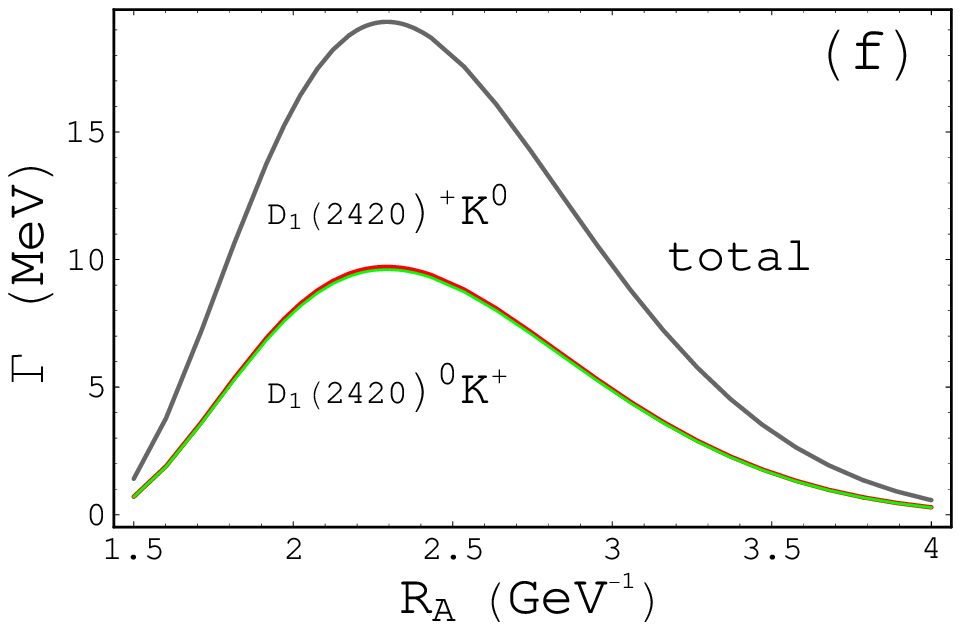}}\\
\scalebox{0.56}{\includegraphics{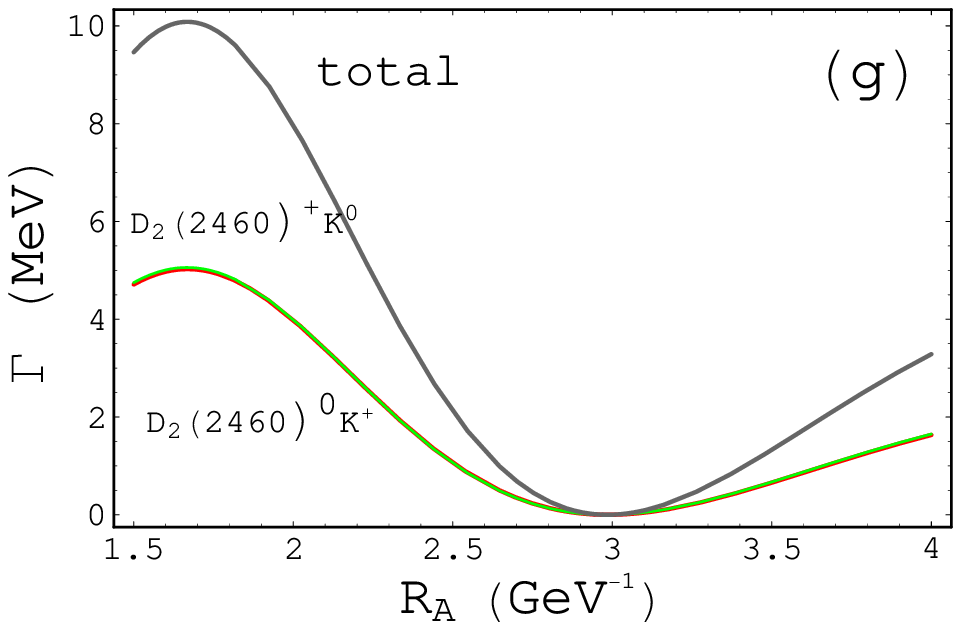}}&
\end{tabular}
\caption{The variation of the strong decays for (a)
$2^+(T^\star)\to 0^-+0^-$, (b) $2^+(T^\star)\to 0^-+1^-$, (c) $2^+(T^\star)\to 1^-+0^-$, (d) $2^+(T^\star)\to 1^-+1^-$,
(e) $2^+(T^\star)\to 1^+(S)+0^-$, (f) $2^+(T^\star)\to 1^+(T)+0^-$ and (g) $2^+(T^\star)\to 2^++0^-$ with the factor $R_A$ of the HO wavefunction of
$2^+(T^\star)$. Here the total partial decay width is labeled by "total" in diagrams. \label{2T}}
\end{figure}
\end{center}

The dependence of the strong decays $2^+(T^\star)\to 0^-+0^-,\,0^-+1^-,\, 1^-+0^-,\,1^-+1^-,\, 1^+(S)+0^-,\, 1^+(T)+0^-,\,2^++0^-$ on the factor $R_A$ ($R$ value of the HO wavefunction of
$2^+(T^\star)$) is given in Fig. \ref{2T}. Here, we take the mass of $2^+(T^\star)$ as 3.157 GeV \cite{Matsuki:2006rz}. When taking $R_{A}=2.8$ GeV$^{-1}$, the total decay width of $2^+(T^\star)$ is 87.9 MeV (see Fig. \ref{compare-1t}), and the partial decay widths (see Fig. \ref{2T}) respectively corresponding to $2^+(T^\star)\to 0^-+0^-,\,0^-+1^-,\, 1^-+0^-,\,1^-+1^-,\, 1^+(S)+0^-,\, 1^+(T)+0^-,\,2^++0^-$ are 15.6 MeV, 0.49 MeV, 7.2 MeV, 49.3 MeV, 1.8 MeV, 13.2 MeV, 0.28 MeV, which show that $1^-+1^-$, $0^-+0^-$, $1^+(T)+0^-$ and $1^-+0^-$ are key decay channels to find $2^+(T^\star)$ charmed-strange meson.

\section{Summary}\label{sec3}

Stimulated by the newly observed charmed-strange meson $D_{sJ}(3040)^+$, we systemically study the two-body strong decays of P-wave charmed-strange mesons with the first radial excitation.

Our numerical results show that $D_{sJ}(3040)^+$ can be categorized as $1^+$ state in $S=(0^+,1^+)$ doublet well, i.e. $D_{sJ}(3040)^+$ is the first radial excitation of $D_{s1}(2460)^+$. We suggest experimentalist to search $D_{sJ}(3040)^+$ by $0^-+1^-$ channel ($D^+K^{*0}$, $D^0 K^{*+}$ and $D_s^+\phi$).

In the past six years, Babar and Belle experiments have made big progress in searching for charmed-strange mesons, which lets us believe that more charmed-strange mesons will be found in future experiment.  If $D_{sJ}(3040)^+$ is the first radial excitation of $D_{s1}(2460)^+$, there must exist three partners of $D_{sJ}(3040)^+$, which are the rest three P-wave charmed-strange mesons with the first radial excitation. In this work, we also study the strong decays of the rest three P-wave charmed-strange mesons with the first radial excitation.
Our numerical result (see the presentation in the subsection of numerical result) will be helpful to instruct future experimental search of the remaining three P-wave charmed-strange mesons with the first radial excitation.

\vspace{1.5cm}

\emph{{\bf Note added.}} When this manuscript was completed, a work of $D_{sJ}(3040)^+$ appeared \cite{Chen:2009zt}. In this work, authors investigated the $D_s$ mesons by a semi-classic flux tube model and explained $D_{sJ}(3040)^+$ as $1^+(j^P=\frac{1}{2}^+)$. In our case, we calculated the strong decays of $D_{sJ}(3040)^+$ with the assignment of the first radial excitation of $D_{s1}(2460)$. By comparing the total decay width of $D_{sJ}(3040)^+$ obtained by the $^3P_0$ model with the Babar data, we conclude that $D_{sJ}(3040)^+$ is the first radial excitation of $D_{s1}(2460)$, which is consistent with the conclusion of the structure of $D_{sJ}(3040)^+$ \cite{Chen:2009zt}.

\section*{Acknowledgement}We are grateful to Prof. Hai-Yang Cheng for suggestive discussion.
This project is support by National Natural Science Foundation of China under Grants 10705001 and A Foundation for the Author of National Excellent Doctoral Dissertation of P.R. China (FANEDD).

\section*{Appendix}

According to the the spatial integral in eq. (\ref{integral}), one can categorize the strong decays of P-wave charmed-strange mesons with the first radial excitation into
two groups: $2P\to 1S+1S$ and $2P\to 1P+1S$.

For the case of $2P\to 1S+1S$, the spatial integral $I^{M_{L_A},m}_{M_{L_B},M_{L_C}}$ is simplified as $I_{m'n'}$ due to $M_{L_B}=M_{L_C}=0$. According to the constraint from eq. (\ref{JB}), we take the direction of $\textbf{K}$ along $z$ axis: $\mathbf{K}=(0,0,|\mathbf{K}|)$. In the following, one presents the result of the spatial integral of $2P\to 1S+1S$ listed in Table \ref{amp}:
\begin{eqnarray}
I_{\pm}&=&I_{1-1}=I_{-11}=2\sqrt{2}\pi^{3/2}\omega_1\Big[\frac{\lambda}{\mathcal{R}^5}+\frac{5}{\mathcal{R}^7}\Big],\\
I_{0}&=&I_{00}=2\sqrt{2}\pi^{3/2}\omega_1\Big[\frac{\Delta(1-\Delta)\lambda |\mathbf{K}|^2}{\mathcal{R}^3}-\frac{\lambda}{\mathcal{R}^5}+\frac{2\Delta(1-2\Delta)|\mathbf{K}|^2}{\mathcal{R}^5}
+\frac{3\Delta(1-\Delta)|\mathbf{K}|^2}{\mathcal{R}^5}-\frac{5}{\mathcal{R}^7}\Big].
\end{eqnarray}

The spatial integrals for $2P\to 1P+1S$, which are involved in the expressions shown in Table \ref{amp}, include
\begin{eqnarray}
I^{00}_{00}&=&-2\sqrt{2}\pi^{3/2}\omega_2\Big\{\frac{\lambda\Delta(\Delta-\mu)(\Delta-1)|\mathbf{K}|^3}{\mathcal{R}^3}+\frac{|\mathbf{K}|}{\mathcal{R}^5}
\Big[\lambda(3\Delta-\mu-1)+2\Delta^2(\Delta-\mu)|\mathbf{K}|^2
+2\Delta^2(\Delta-1)|\mathbf{K}|^2\nonumber\\&&+2
\Delta(\Delta-\mu)(\Delta-1)|\mathbf{K}|^2+3\Delta(\Delta-\mu)(\Delta-1)|\mathbf{K}|^2\Big]
+\frac{5(3\Delta-\mu-1)|\mathbf{K}|}{\mathcal{R}^7}+\frac{6\Delta|\mathbf{K}|}{\mathcal{R}^7}\Big\},\\
I^{01}_{10}&=&I^{0-1}_{-10}=-2\sqrt{2}\pi^{3/2}\omega_2\Delta |\mathbf{K}|\Big[\frac{\lambda}{\mathcal{R}^5}+\frac{7}{\mathcal{R}^7}\Big],\\
I^{10}_{10}&=&I^{-10}_{-10}=-2\sqrt{2}\pi^{3/2}\omega_2\Big[\frac{\lambda(\Delta-1)|\mathbf{K}|}{\mathcal{R}^5}+\frac{(7\Delta-5)|\mathbf{K}|}{\mathcal{R}^7}\Big],\\
I^{1-1}_{00}&=&I^{-11}_{00}=2\sqrt{2}\pi^{3/2}\omega_2\Big[\frac{\lambda(\Delta-\mu)|\mathbf{K}|}{\mathcal{R}^5}+\frac{(7\Delta-5\mu)|\mathbf{K}|}{\mathcal{R}^7}\Big].\\
\end{eqnarray}

Here,
\begin{eqnarray}
\mathcal{R}&=&\sqrt{R_A^2+R_B^2+R_C^2},\quad \mu=\frac{m_1}{m_1+m_3},\quad \nu=\frac{m_2}{m_2+m_4},\nonumber\\
\eta&=&\frac{R_B^2\mu+R_C^2\nu}{\sqrt{R_A^2+R_B^2+R_C^2}},\quad \xi^2=R_B^2\mu^2+R_C^2\nu^2-\eta^2,\quad\Delta=\frac{\eta}{\mathcal{R}},\quad\nonumber\\
\lambda&=&-\frac{5-2R_A^2\Delta^2|\mathbf{K}|^2}{2R_A^2},\nonumber\\
\omega_1&=&-\frac{3iR_A^3(R_A R_B R_C)^{3/2}}{\sqrt{15}\pi^{11/4}}\exp\Big(-\frac{1}{2}\xi^2|\mathbf{K}|^2\Big),\nonumber\\
\omega_2&=&\frac{\sqrt{6}R_A^2(R_A R_B R_C)^{5/2}}{\sqrt{5}\pi^{11/4}R_{C}}\exp\Big(-\frac{1}{2}\xi^2|\mathbf{K}|^2\Big).\nonumber
\end{eqnarray}

\end{document}